\documentclass[11pt,a4paper,openbib]{article}
\usepackage{amssymb}
\usepackage{graphicx}

\newif\ifpdf
\ifx\pdfoutput\undefined
  \pdffalse 
\else
  \pdfoutput=1 
  \pdftrue
\fi
\ifpdf
  \DeclareGraphicsExtensions{.pdf,.png,.jpg,.mps}
\else
  \DeclareGraphicsExtensions{.ps,.eps}
\fi

\if@twoside
\else
\fi
\def\vac#1{|{#1}\rangle}

\def\avac#1{\langle{#1}|}

\def\vev#1{\langle{#1}\rangle}

\def\Vev#1{\left\langle{#1}\right\rangle}
\def\vvev#1{\langle\!\langle{#1}\rangle\!\rangle}

\def\Vvev#1{\left\langle\kern-5pt\left\langle{%
  #1}\right\rangle\kern-5pt\right\rangle}
\def\Vvevs#1{\left\langle\kern-4pt\left\langle{%
  #1}\right\rangle\kern-4pt\right\rangle}
\def\Vvevss#1{\left\langle\kern-3pt\left\langle{%
  #1}\right\rangle\kern-3pt\right\rangle}
\def\vec#1{{\rm\bf#1}}
\def\nop#1{\mbox{:$#1$:}}
\def\cal{\mathcal}
\def\IM{\Im\mathfrak{m}}
\def\RE{\Re\mathfrak{e}}
\def\bra{\langle}
\def\ket{\rangle}

\begin{document}

\title{%
{\scshape Logarithmic Conformal Field Theory}\\ 
{\scshape-- or --}\\ 
{\scshape How to Compute a Torus Amplitude on the Sphere}\\
\begin{picture}(0,0)(0,0)
  \put(141,137){{\small{\tt hep-th/0407003}}}
  \put(-215,137){{\small{\tt BONN-TH-2004-03}}}
  \put(-215,135){\line(1,0){430}}
\end{picture}
}

\author{\phantom{Quark}\\
Mic$\hbar$ael A.I.~Flohr\footnote{Work 
supported by the European Union network 
HPRN-CT-2002-00325 and in part by the string theory 
network (SPP no.\ 1096), Fl 259/2-2, of the 
Deutsche Forschungsgemeinschaft.}\\[1em]
\addtocounter{footnote}{5}%
%
%
{\slshape Physics Institute\/}\footnote{On leave of absence from
Institute for Theoretical Physics, Univesity of Hannover, Appelstr.~2,
D-30167 Hannover, Germany.}\\
{\slshape University of Bonn}\\ 
{\slshape Nussallee 12, D-53115 Bonn, Germany}\\ 
{\tt flohr@th.physik.uni-bonn.de}\\
}


\maketitle

\abstract{

\medskip\noindent%
We review some aspects of logarithmic conformal field theories which might
shed some light on the geometrical meaning of logarithmic operators.
We consider an approach, put forward by V.~Knizhnik, where computation of
correlation functions on higher genus Riemann surfaces can be replaced
by computations on the sphere under certain circumstances. We show that
this proposal naturally leads to logarithmic conformal field theories, when
the additional vertex operator insertions, which simulate the branch
points of a ramified covering of the sphere, are viewed as dynamical 
objects in 
\phantom{xxxxxxxxxxxxx}\hfill the theory.\hfill \phantom{xxxxxxxxxxxxx}

\medskip\noindent%
We study the Seiberg-Witten solution of supersymmetric low energy effective 
field theory as an example where physically interesting quantities, the
periods of a meromorphic one-form, can effectively computed within this
conformal field theory setting. We comment on the relation between
correlation functions computed on the plane, but with insertions 
of twist fields, and torus vacuum 
\phantom{xxxxxxxxxxxxx}\hfill amplitudes.\hfill \phantom{xxxxxxxxxxxxx}}

\bigskip\bigskip\centerline{{\slshape In memoriam Ian Kogan.}}
\newpage
\tableofcontents
\newpage

\section{Introduction}

Riemann surfaces belong to this special class of beautiful geometrical
objects, whose members on one hand enjoy a sheer inexhaustible richness
of mathematical structure, and one the other hand show up surprisingly
(or even suspiciously?) often in the attempts of theoretical physicists to
uncover fundamental patterns of what we call reality.

One of the more prominent appearances of Riemann surfaces in theoretical
physics is, beyond any doubt, the worldsheet swept out by a string. 
Parameterization invariance in string theory then led us to conformal field
theories (CFT) as the natural inhabitants of these worldsheets -- and
hence Riemann surfaces. A particular useful species of them, possessing a
finite operator algebra, are the so-called rational conformal field
theories (RCFTs), a concept first introduced with the minimal
models of Belavin, Polyakov and Zamolodchikov \cite{BPZ83} more than
15 years ago. Since then RCFTs established themselves as
a main tool in modern theoretical physics, with deep relations to a
multitude of mathematical fields.

String theories have changed much since their childhood, and their modern 
counterparts such as $M$-theory, unifying all the former theories, are now 
roamed by lots of other beings of higher dimensionality, branes and their
worldvolumes. However, new developments in 4-dimensional effective field
theories gave Riemann surfaces a new task. They now serve as the moduli
spaces of the vacua of exactly solvable effective field theories, namely
supersymmetric Yang-Mills theories. Strikingly, these effective
field theories can be derived as certain low-energy limits of, e.g.\ 
type IIB string theory or $M$-theory. The fact that the moduli spaces turn
out to be Riemann surfaces then is rooted in special properties of
higher-dimensional geometrical objects such as Calabi-Yau compactifications 
in type IIB or the worldvolume of the 5-brane in $M$.

The main goal of this paper, which is devoted to the memory of Ian Kogan, 
is to test a more recently
evolved species of CFTs for its suitability for theoretical physics --
the so-called logarithmic conformal field theories (LCFTs). Although RCFTs
nicely encode the additional degrees of freedom of string theories on
their respective worldsheets, they completely fail to describe the physics
in the moduli spaces of effective super-YM theories. This is simply due to
the fact that RCFT correlation functions never ever produce logarithms,
and hence cannot reproduce the logarithmic dependencies of the periods.
Nonetheless, LCFTs do, and therefore we propose LCFTs as the
natural candidates to encode the BPS spectrum of effective low-energy
field theories. 

The ideas we collect in this paper have partially been
published in my earlier work \cite{Flo98}. Many discussions with Ian flow into
that work. We both were very much intrigued by an old paper, written by
Knizhnik. Knizhnik uses ghost system CFTs to compute string amplitudes up to
two loops, i.e.\ genus two Riemann surfaces. The trick, Knizhnik uses, is to
represent the complicated Riemann surfaces as a branched covering of the
Riemann sphere and then simulate the effect of the branch points by special
vertex operators. What made us so excited about Knizhnik's work is that
he notes that certain correlators involving these branch point vertex
operators may exhibit a logarithmic divergence. To my knowledge, this is
the earliest mentioning of this possibility in the literature. Knizhnik
discusses this issue only briefly, dismissing the possibility as, in his
context correctly, physically irrelevant. Therefore, I find it appropriate
to devote my paper to this common interest of both of us.

Logarithmic conformal
field theories, first encountered and shown to be consistent in
\cite{Gur93}, are not just a peculiarity but merely a generalization of
ordinary two-dimensional CFTs with broad and growing applications. 
One may well say that LCFTs contain ordinary rational conformal field
theories as just the subset of theories free of logarithmic
correlation functions. However, logarithmic divergences are sometimes
quite physical, and so there is an increasing interest in these logarithmic
conformal field theories. 

The body of LCFT literature is by now too large to be listed here in
completeness. 
We will here only note a few of the applications, confining ourselves to works
which existed at the time when Ian and I discussed LCFT in connection with
Seiberg-Witten theory. These are works on topics such as (multi-)critical 
polymers and percolation in two dimensions 
\cite{DuSa87,Flo95,Iva98,Sal92,Wat96}, two-dimensional turbulence and
magneto-hydrodynamics \cite{Flo96a,RaRo95,SkTh98}, the quantum Hall effect
\cite{Flo96,GFN97,MaSe96,WeWu94}, gravitational dressing 
\cite{BiKo95,KoLe97,KLS97}, disorder and localization effects
\cite{CKT95,CTT98,KMT96,MaSe96} as well as the recoil problem of strings 
and $D$-branes \cite{BCFPR96,EMN96,KoMa95,KMW96,PeTa96} (see
\cite{Nak04} and references therein for more recent work) and target-space 
symmetries in string theory in general \cite{KoMa95}.

There is also a growing body of literature on LCFT in general, where
in particular all the powerful structures of RCFTs are either transfered
or generalized to the logarithmic case. For example, LCFTs appear in the 
WZNW model on the supergroup $GL(1,1)$ \cite{RoSa92}, the $c_{p,1}$ models 
(as well as non-minimal $c_{p,q}$ models)
\cite{Flo95,GaKa96,Gur93,Kau95,Roh96}, gravitationally dressed conformal
field theories \cite{BiKo95}, WZNW models at level $0$ \cite{KoMa95,CKLT97} or
fractional \cite{Gaberdiel:2001a} or negative \cite{Kogan:2001} level,
and critical disordered models \cite{CKT95,CTT98,GuLu99,MaSe96}. The references
on the $c_{p,1}$ models also contain considerable information on general
questions such as characters, fusion rules, partition functions and
null vectors (see also \cite{Flo97}), 
while logarithmic correlation functions were considered in 
general in \cite{GhKa97,KoLe97,KLS97,MRS00,RAK96,ShRa96}, 
see also \cite{RaRo97} 
about consequences for Zamolodchikov's $C$-theorem. Generic construction
schemes for LCFTs were discussed in \cite{Fjelstad}.

For more recent works and a more detailed exposition on the state of the
art of LCFT, the reader is referred to the reviews \cite{Flohr:2001,
Gaberdiel:2001} and references therein.

The aim of this paper is to establish LCFTs as natural
inhabitants of Riemann surfaces, providing a geometrical foundation for
their special properties which distinguish them from ordinary CFTs.
Furthermore, this is done with precisely such physical applications in mind, 
where Riemann surfaces play an important r\^ole, but ordinary RCFTs fail
in effectively computing the desired information. Therefore, this paper will
study LCFTs in the framework of certain classes of Riemann surfaces
which, in particular, appear as the moduli spaces of vacua of exactly solvable
supersymmetric Yang-Mills theories. 

We will focus on the simplest case of non-trivial Riemann
surfaces, the hyperelliptic ones. Such complex curves describe the moduli 
spaces of vacua of Seiberg-Witten models, i.e.\ exact solutions of low-energy
$N$=2 supersymmetric Yang-Mills theories. Much of what is collected here
can easily be generalized to
the case of general Riemann surfaces with a global $\mathbb{Z}_n$ symmetry, 
which serve, for instance, as the moduli spaces of arbitrary four-dimensional
supersymmetric gauge theories with reductive gauge groups, whose solutions
have been found by Witten via $M$-theory. 

The particular choice of our application, namely low-energy effective
field theories, is motivated twofold: Firstly, these theories are of
particular interest for the question of duality and for the formulation
of high-energy theories (string theory, $M$-theory, etc.) which are
phenomenological promising. Secondly, the so-called Seiberg-Witten models
are good examples where one knows that a certain geometrical object, namely
the hyperelliptic curve, encodes entirely all information about the theory,
but where it can still be quite difficult to extract this information
explicitly, meaning here to calculate the periods of certain forms.

Furthermore, this application illuminates the geometry behind logarithmic
CFT. It is well known that vertex operators of worldsheet CFTs in string 
theory describe the equivalent of Feynman graphs with outer legs by 
simulating their effect on a Riemann surface as punctures. Now, in the new
setting of moduli spaces of low-energy effective field theories, {\em pairs
of\/} vertex operators describe the insertion of additional handles to a
Riemann surface, simulating the resulting branch cuts. So, in much the same
way as a smooth but infinitely long stretched tube attached to an otherwise
closed worldsheet, standing for an external state, is replaced by a
puncture with an appropriate vertex operator, so is a smooth additional
handle, standing for an intersecting 4-brane on the 5-brane worldvolume (after
switching on the 11-th dimension in $M$-theory), replaced by branch cuts
with appropriate vertex operators at its endpoints.

The paper is structured as follows:
Hence, section II reviews in general, how vertex operators can be
used to describe the behavior of a $n$-ramified branch cut before 
concentrating on the hyperelliptic case and identifying the correct
CFTs for the description of $j$-differentials. It is shown that in the
case of our focus, this is the rational LCFT with $c=c_{2,1}=-2$. 
Finally, arbitrary Abelian differential 1-forms on 
hyperelliptic curves are defined in terms of 
conformal blocks of suitable vertex operators. 

Section III gives a brief exposition of the Seiberg-Witten solutions of
$N$=2 supersymmetric four-dimensional Yang-Mills theories before going
on with the calculation of the periods of the meromorphic Seiberg-Witten
differential. The case of gauge group $SU(2)$ is performed in some
detail to show the naturality of the CFT approach and its potential
to considerably simplify such computations. 

The IV.\ section will then concentrate on the case of the torus in order
to make the relation of Knizhnik's approach to objects, which one can
compute directly on the torus, very explicit. For example, we will show there, 
that torus vacuum amplitudes can be evaluated directly and also as
four-point functions on the sphere yielding exactly the same result. We
also comment on the possibility to generalize this approach to other
conformal field theories.

After the obligatory conclusions, an appendix presents all important
structure constants of the $c=-2$ theory, which are needed when $n$-point
correlation functions are expressed in terms of 4-point functions, as well
as some information on multiple hypergeometric functions.

\section{The approach of Knizhnik}

The idea to compute integrals of differential forms on Riemann surfaces 
with the
help of conformal field theory is actually not new. Some path-breaking
papers on this idea are \cite{BeRa86,DFMS86,Kni86,Zam86}. However, as has
been shown \cite{Kro02}, 
the conformal field theories of twist and spin fields
used in these earlier works are indeed very special logarithmic conformal 
field theories. As a consequence of this, we are now -- after the advent of
logarithmic conformal field theory -- in a better position: Logarithmic
conformal field theories possess a structure very close to rationality
\cite{Flo95,GaKa96} which allows us to make use of all the powerful tools 
available in rational conformal field theory. Not only can
explicit calculations of period integrals be performed within the setting
of degenerate conformal field theories \`a la BPZ \cite{BPZ83}, but also the
physical interpretation of such periods, as in the Seiberg-Witten models,
becomes more transparent. For example, the fusion rules yield a very simple
and imaginative way to find the points in the moduli space of a Seiberg-Witten
model, where certain states become massless. Moreover, the description within 
the logarithmic conformal field theory setting contains some surprising new
structures not apparent in the older approach, especially the feature of
Jordan-cell highest weight representations and the fact that the zero mode
$L_0$ of the stress-energy tensor cannot be diagonalized.

The basic idea of the work of Knizhnik is that a given compact Riemann surface
can always be represented as a $n$-fold ramified covering of the complex
plane or, more precisely, its compactified version, the Riemann sphere
$\mathbb{CP}^1$, for some number $n$. Thus, instead of computing a
correlation function on the non-trivial Riemann surface, one could attempt
to compute a correlator on the complex sphere with additional insertions of
suitable operators which precisely have the effect of the branch points.
Knizhnik explicitly constructed such vertex operators for the special case
of the so-called ghost systems. These are particular conformal field
theories of a free pair of anti-commuting fields $b$ and $c$ with conformal
scaling dimension $j$ and $1-j$ respectively. We briefly review his results
here.

\subsection{The Conformal Field Theory of $j$-Differentials}
An arbitrary Riemann surface $X$ can be represented as a branched covering of
$\mathbb{CP}^1$ where the covering map is denoted by $Z$. 
The metric on $X$ can be
chosen as $g_{zz}=g_{\bar z\bar z}=0$, $g_{z\bar z}=1$. Under this choice,
it was shown in \cite{Kni86} that each branch point $e_i$ corresponds to a 
particular primary field $\Phi_i(e_i)$. Usually, we will work with Riemann
surfaces where the ramification numbers of all branch points are equal, say
$n$, which means that $X$ has a global $\mathbb{Z}_n$ symmetry and can be 
defined by an equation
\begin{equation}\label{eq:Zn}
   y^n = \prod_{k=1}^{nm}(Z-e_k)\,,\ \ \ \ g=(n-1)(\frac{nm}{2}-1)\,,
\end{equation}
in $\mathbb{C}^2=(y,Z)$. The hyperelliptic case, which will be the main scope of
the present paper, corresponds to $n=2$, i.e.\ $\mathbb{Z}_2$ symmetry.

We now briefly review the construction of $j$-differentials and branch
points in terms of primary fields. Near a branch point $a$ of order $n$, the
covering map $Z$ takes locally the form $Z(y) = a + y^n$ yielding $n$ sheets
of $X$ via the inverse map $y(Z) = (Z-a)^{1/n}$, which we enumerate by
$\ell=0,\ldots,n-1$. Thus, moving a point around $Z=a$, its inverse image
moves $\ell\mapsto\ell+1$ mod $n$. We denote this analytic continuation
operation by $\hat{\pi}_a$. For each sheet, we consider a pair 
$\phi^{(j),\ell}, \phi^{(1-j),\ell}$ of anticommuting fields of spin 
$j$ and $1-j$ respectively, and the action
\begin{equation}
  S^{(\ell)} = \int\,\phi^{(j),\ell}\,\bar{\partial}\phi^{(1-j),\ell}\,
  {\rm d}^2Z\,.
\end{equation}
In order that the anticommuting fields are local chiral fields, 
the spin $j$ must be (half-) integer.
It is well known that the stress energy tensor takes the form
\begin{equation}
   T^{(\ell)} = -j\phi^{(j),\ell}\,\partial\phi^{(1-j),\ell}
              + (j-1)\phi^{(1-j),\ell}\,\partial\phi^{(j),\ell}
\end{equation}
giving rise to a central extension $c=c_j\equiv-2(6j^2-6j+1)$. The point is
that under a conformal transformation $Z\mapsto Z'(Z)$, $\bar Z\mapsto 
\overline{Z'(Z)}$, these fields transform as $j$- and $(1-j)$-differentials
respectively:
\begin{equation}
  \phi^{(j),\ell}(Z',\bar Z')\left(\frac{{\rm d}Z'}{{\rm d}Z}\right)^j =
  \phi^{(j),\ell}(Z,\bar Z)\,, 
\end{equation}
and analogously for $\phi^{(1-j),\ell}$.
Here we assume that the operator product expansion (OPE) be normalized as 
\begin{equation}\label{eq:openorm}
  \phi^{(j),\ell}(Z')\phi^{(1-j),\ell}(Z)\simeq\mathbb{I}\,(Z'-Z)^{-1} + 
  \textrm{regular\ terms}
\end{equation}
with $\mathbb{I}$ denoting the identity operator.
Essentially, these fields are trivializing sections of $K^j$ on the Riemann 
surface $X$, where $K$ denotes the canonical line bundle. The central 
extension of the Virasoro algebra can than be expressed as
$c = 2c_1[\det\bar{\partial}_{(j)}]$, i.e.\ by the first Chern class of the
determinant of the holomorphic derivative acting on $j$-differentials.

The boundary conditions of these fields with respect to the operator
$\hat{\pi}_a$ are for arbitrary $j\in\mathbb{Z}/2$
\begin{equation}
  \hat{\pi}_a\phi^{(j),\ell}(Z)=(-)^{2j}\phi^{(j),\ell+1\,{\rm mod}\,n}(Z)
  \,.
\end{equation}
In the vicinity of the branch point we can diagonalize $\hat{\pi}_a$ by
choosing another basis for the $j$-differentials via a discrete Fourier
transform,
\begin{equation}
  \phi^{(j)}_k = 
  \sum_{\ell=0}^{n-1}{\rm e}^{-2\pi{\rm i}(k+j-nj)\ell/n}\phi^{(j),\ell}\,,
\end{equation}
such that $\hat{\pi}_a\phi^{(j)}_k = {\rm e}^{2\pi{\rm i}(k+j-nj)/n}
\phi^{(j)}_k$.
As a consequence, we can define currents $J_k$ which are single-valued
functions of $Z$ in the vicinity of the branch point $a$. They are
given by $J_k = \nop{\phi^{(j)}_k\phi^{(1-j)}_{n-1-k}}$ and are chiral,
$\bar{\partial}J_k=0$.
These currents give rise to charges, and it turns out that a branch point
carries the charges
\begin{equation}\label{eq:charge}
  q_k = \frac{(n-1)j-k}{n}\,.
\end{equation}
We may now perform a bosonization procedure expressing all fields with the
help of $n$ analytic scalar fields $\varphi_k$, $k=0,\ldots,n-1$, normalized
as $\vev{\varphi_k(Z)\varphi_l(Z')}=-\delta_{kl}\log(Z-Z')$. We find that,
for instance, $\phi^{(j)}_k = \nop{{\rm e}^{-i\varphi_k}}$, 
$\phi^{(1-j)}_{n-1-k} = \nop{{\rm e}^{i\varphi_k}}$ and $J_k 
= {\rm i}\partial\phi_k$.
One might attempt to diagonalize the stress-energy tensor too, obtaining
\begin{equation}
  T_k = -j\phi^{(j)}_k\partial\phi^{(1-j)}_{n-1-k} 
  + (1-j)\phi^{(1-j)}_{n-1-k}\partial\phi^{(j)}_k = 
  \frac{1}{2}\left(\nop{J_kJ_k} - (2j-1)\partial J_k\right)\,.
\end{equation}
However, we will see later that under certain circumstances the zero mode of 
the stress energy tensor cannot be diagonalized.
It follows that a branch point $a$ is represented by a vertex operator
\begin{equation}
  V_{\vec{q}}(a) = \nop{\exp({\rm i}\vec{q\mbox{{\boldmath$\varphi$}}}(a))}
  \,,\ \ \ \
  \vec{q\mbox{{\boldmath $\varphi$}}} = \sum_k q_k\varphi_k\,,
\end{equation}
having conformal scaling dimension $h = h(\vec{q}) = \sum_k h_k$ with the
$h_k$ given by $h_k=\frac{1}{2}(q_k^2-(2j-1)q_k)$.
This corresponds to a Coulomb gas like construction of our conformal field
theory (CFT) with a background charge $-2\alpha_0$ given by 
$c_j=1-12\alpha_0^2$, i.e.\ $2\alpha_0 = 2j-1$.
The screening charges for the resulting degenerate model are
$\alpha_{\pm} = \alpha_0 \pm \sqrt{\alpha_0^2+2}$. Actually, the full
CFT consists of $n$ copies, one for each sheet, with
total central charge $c=nc_j$. 

In the case that $X$ has a global $\mathbb{Z}_n$ symmetry (\ref{eq:Zn}), all
operators $\hat{\pi}_{e_k}$ can be diagonalized simultaneously, meaning that
we can define our CFT globally, i.e.\ everywhere on
the surface. Furthermore, the CFT is then just an $n$-fold
tensor product of the simple CFT of two anticommuting analytic fields of
spin $j$ and $(1-j)$ respectively. It is therefore sufficient to consider
only one of these copies for most of our issues. Hence, we often will drop
the indices $\ell$ or $k$. Clearly, according to (\ref{eq:openorm}),
$\phi^{(1-j)}$ is the conjugate of $\phi^{(j)}$
with respect to the canonical scalar product $\langle\cdot,\cdot\rangle =
\frac{1}{2\pi{\rm i}}\oint{\rm d}z$. We also note that the OPE
of the basic vertex operators has as its first term
\begin{equation}
  V_{\vec{q}}(a)V_{\vec{q'}}(a') \simeq (a-a')^{\vec{q}\cdot\vec{q'}}
  V_{\vec{q}+\vec{q'}}(a') + \ldots\ .
\end{equation}
Usually, this is the leading term, since the order of the singularities on the
right hand side is given by $h(\vec{q''})-h(\vec{q})-h(\vec{q'})$ and has
a minimum for $\vec{q''} = \vec{q}+\vec{q'}$. Hence, correlation functions
of vertex operators are simply evaluated as
\begin{equation}
  \Vev{\prod_{j=1}^{N}V_{\vec{q}_j}(z_j)} =
  \prod_{i<j}(z_i-z_j)^{\vec{q}_i\cdot\vec{q}_j}
\end{equation}
with the charge balance condition $\sum_{i=1}^{N}\vec{q}_i=2\alpha_0\vec{1}$.

What we have done is essentially defining a map from Riemann surfaces $X$ to
the Fock space ${\cal F}$ of semi-infinite forms on the Hilbert space 
${\cal H} = L^2(S^1)$. This
map is called the string map. The Fock space is graded with respect to the
charges $\vec{q}$, ${\cal F} = \bigoplus_{\vec{q}}{\cal F}_{\vec q}$, and 
admits a representation of the Virasoro algebra as the
central extension of ${\it Diff}(S^1)$. We have been a bit sloppy with the
definition of vertex operators, since these are actually maps
$V_{\vec{q},\vec{p}}^{\vec{r}}(\cdot,Z) : {\cal F}_{\vec{q}} \mapsto
{\rm Hom}({\cal F}_{\vec{p}},{\cal F}_{\vec{r}})$, called the screened
chiral vertex operators. The screening refers to the fact that the 
3-point function $\langle h_{\vec{r}}|Q_-^{\vec{s}_-}Q_+^{\vec{s}_+}
V_{\vec{q}}(1)|h_{\vec{p}}\rangle$ is only non-zero, if $\vec{p}+\vec{q}
+\vec{r}+\alpha_-\vec{s}_-+\alpha_+\vec{s}_+=\vec{0}$. Thus,
$V_{\vec{q},\vec{p}}^{\vec{r}}(\cdot,Z)$ denotes a vertex operator with
the appropriate numbers of screening charges attached. The chiral primary fields
are certain linear combinations of these chiral vertex operators with
coefficients determined by locality and crossing symmetry of the primary
fields of the complete CFT, combined from its left and right chiral half.
In the following, we will denote by $\Phi_{\vec{q}}$ the primary
fields, which are given as
\begin{equation}\label{eq:decomp}
  \Phi_{\vec{q}}(Z,\bar{Z}) = \sum_{\vec{p},\vec{r}}\,
  {\cal D}_{\vec{q},\vec{p}}^{\vec{r}}\,
  V_{\vec{q},\vec{p}}^{\vec{r}}(\cdot,Z)
  V_{\bar{\vec{q}},\bar{\vec{p}}}^{\bar{\vec{r}}}(\cdot,\bar{Z})\,.
\end{equation}
We will determine the coefficients ${\cal D}_{\vec{q},\vec{p}}^{\vec{r}}$
for the case most important to us, $j=1$, in the Appendix A. These
coefficients are closely related to the structure constants of the OPE,
which can also be found in this Appendix. Let us further remark that,
as evident from the construction, the CFT for the $j$-differentials is
dual equivalent to the CFT for the $(1-j)$-differentials.

\subsection{The Rational Logarithmic Conformal Field Theory with $c=-2$}

In this paper, we concentrate on the $\mathbb{Z}_2$ symmetric case of Riemann
surfaces, i.e.\ the case of hyperelliptic curves. Such a curve of genus $g$
can be represented by a double covering of the complex plane
\begin{equation}\label{eq:curve}
  y^2 = \prod_{k=1}^{2g+2}(z-e_k)\,,\ \ \ \ \sum_{k=1}^{2g+2}e_k = 0\,,
\end{equation}
where we assume that $e_k\neq e_l$ for $k\neq l$ and that infinity is not
a branch point. However, as we will see later, our formalism naturally
incorporates the case of a degenerating curve (where two branch points flow
together) as well as the case that infinity is a branch point.
If we have made a choice for the branch cuts, we will display this by
writing
\begin{equation}
  y^2 = \prod_{k=1}^{g+1}(z-e_k^-)(z-e_k^+)\,,\ \ \ \ \sum_{k=1}^{g+1}
  e_k^{\pm} = 0\,,
\end{equation}
with the $g+1$ branch cuts running between $e_k^-$ and $e_k^+$.
Moreover, we mainly want to evaluate holomorphic or meromorphic
1-differentials, e.g.\ determine their periods. That means that we are
interested in the case $j=1$. Then, the conformal field theory
of twist fields appropriate for this case can be constructed out of two
anticommuting fields of spin 1 and 0, and has central charge $c=-2$.
The branch points are represented by primary fields of conformal
weight $h=-1/8$. This construction follows from first principles as shown in
\cite{Kni86} and sketched in the preceding subsection. However, any conformal 
field theory with $c=-2$ and a primary
field of conformal weight $h=-1/8$ is logarithmic \cite{Gur93}. 

Hence, we can associate a branch point with certain primary fields 
$\Phi_{q_k}$ having charges $q_k$ and conformal weights $h_k =
\frac{1}{2}(q_k^2 - q_k)$. It turns out that $\Phi_{q_0}=\Phi_{1/2}$ is a 
twist field, and $\Phi_{q_1}=\Phi_0\equiv\mathbb{I}$ is nothing else than 
the identity. Of course, the correct fields of the full
conformal field theory for both sheets are the sums of the fields for each
sheet, $\Phi_{\vec{q}}=\Phi_{(q_0,q_1)}$. Thus, the full conformal field 
theory has $c=2(-2)=-4$ with two
identical versions for each field and is therefore a tensor product of 
two $c=-2$ theories. Since for the branch points $\vec{q}=(q,0)$ is trivial
on the second sheet,
it follows that it is sufficient to consider only one copy of the $c=-2$
theories.\footnote{\,This is a peculiarity of the
$\mathbb{Z}_2$ case. The general case is more involved.}

Let us now have a closer look at this logarithmic conformal field theory 
with $c=c_{2,1}=-2$, which is somehow the ``first'' model in the minimal
series. Although the conformal grid is empty, one can consider the fields
on its boundary. As was shown in \cite{Flo95,GaKa96}, there is a finite
set of primary fields which is closed under the fusion rules. All $c_{p,1}$
models, $p\in\mathbb{Z}_+$, $p>2$, are rational logarithmic conformal field 
theories whose conformal grid can formally be obtained by considering them as
$c_{3p,3}$ models. In our case, we have five primary fields
$\{\mathbb{I},\mu,P,\sigma,J\}$ of conformal weights 
$h\in\{0,-\frac18,0,\frac38,1\}$
respectively. Note that there are {\em two\/} fields of the {\em same\/}
conformal scaling dimension $h=0$. The field $\mathbb{I}$ is just the identity,
while the field $P$ is closely related to the so-called puncture operator
\cite{KoLe97}. The charges of these fields are then 
$\vec{q}\in\{(0,0),(\frac12,0),(1,0),(-\frac12,0),(-1,0)\}$
respectively, and the two screening currents carry charges 
$\alpha_+=2,\alpha_-=-1$. Note that the
screening current $J_-$ is identical to the primary field $J$, which is
a special feature of logarithmic conformal field theory, where the
screening charges become themselves local chiral primary fields \cite{Flo95}.
The operator product expansion of any two fields $\Phi_{\vec{q}},
\Phi_{\vec{q'}}$ with 
charges $\vec{q}=(q,0),\vec{q'}=(q',0)$ has as its leading order
\begin{equation}\label{eq:opeq}
    \Phi_\vec{q}(z)\Phi_{\vec{q'}}(w) \simeq (z-w)^{qq'}C_{q,q'}^{q+q'}
    \Phi_{\vec{q}+\vec{q'}}(w) + \ldots
\end{equation}
with $C_{q,q'}^{q+q'}$ denoting the structure constants of the operator
algebra.
We will freely use both notations of the primary fields, either 
$\Phi_\vec{q}(z)$ with $\vec{q}=(q,0)$ the appropriate charge, or 
$\mathbb{I}(z),\mu(z)$ etc. The important thing
to remember is that for $j=1$, the resulting CFT is a {\em degenerate\/}
model. The subset of degenerate primary fields yields a closed operator
algebra, and the corresponding charged Fock spaces admit well-defined
screening charges. The admissible charges are given as
\begin{equation}\label{eq:admissible}
    q_{r,s} = {\textstyle\frac{1}{2}}\left((1-r)\alpha_+ + (1-s)\alpha_-
    \right) 
\end{equation}
with $r,s$ non-negative integers. Since the $c=-2$ model is even rational,
the conformal grid truncates, meaning $1\leq r\leq 2$, $1\leq s \leq 5$, 
which yields precisely the conformal weights
\begin{equation}
  h_{r,s} = \frac{1}{2}(q_{r,s}^2 - q_{r,s})
\end{equation}
given above. According to this formula, we may also express the OPEs
(\ref{eq:opeq}) in terms of conformal weights instead of charges, namely
\begin{eqnarray}
  & & \Phi_{r,s}(z)\Phi_{r'\!,s'}(w) \simeq \nonumber\\
  & & (z-w)^{h_{r+r'-1,s+s'-1}-h_{r,s}-h_{r'\!,s'}}
  C_{h_{r,s},h_{r'\!,s'}}^{h_{r+r'-1,s+s'-1}}
  \Phi_{r+r'-1,s+s'-1}(w) + \ldots\phantom{xxx}
\end{eqnarray}
We summarize all important data of the $c=-2$ rational LCFT in the following
table, where $q^*$ denotes the charge of the conjugate field, $q^* = 
2\alpha_0 - q$, with respect to the standard (contravariant) pairing 
$\langle\Phi_q|\Phi_p\rangle$ of states $|\Phi_p\rangle\in{\cal F}_p$,
$\langle\Phi_q|\in{\cal F}^*_q\simeq{\cal F}_{q^*}$.
By definition, $h(q^*)=h(q)$, and the background charge 
for $c=-2$ is $2\alpha_0 = 1$. Keeping in mind that the
fields $\sigma$ and $J$ are spin doublets results in a different choice
for $q$ for them so that we have:
{\small $$\begin{array}{|c||c|c|c|c|c|c|c|}\hline
  \Phi_q          & \mathbb{I} & \mu  &  P & \sigma &  J & J_- & J_+ \\ \hline
                                                                 \hline
  (r,s)           & {\scriptstyle (1,1)=(2,5)} 
                  & {\scriptstyle (1,2)=(2,4)} 
                  & {\scriptstyle (1,3)=(2,3)} 
                  & {\scriptstyle (2,2)=(1,4)} 
                  & {\scriptstyle (2,1)=(1,5)} 
                  & {\scriptstyle (1,-1)=(2,7)} 
                  & {\scriptstyle (1,5)=(2,1)} \\ \hline
  h_{r,s}         &  0  & -1/8 &  0 &  3/8   &  1 &  1  &  1  \\ \hline
  q_{r,s}         &  0  &  1/2 &  1 & -1/2   & -1 & -1  &  2  \\ \hline
  q^*=q_{3-r,6-s} &  1  &  1/2 &  0 &  3/2   &  2 &  2  & -1  \\ \hline
  \alpha_-q       &  0  & -1/2 & -1 &  1/2   &  1 &  1  & -2  \\ \hline
  \alpha_-q^*     & -1  & -1/2 &  0 & -3/2   & -2 & -2  &  1  \\ \hline
  \alpha_+q       &  0  &  1   &  2 & -1     & -2 & -2  &  4  \\ \hline
  \alpha_+q^*     &  2  &  1   &  0 &  3     &  4 &  4  & -2  \\ \hline
  \end{array}
$$}

\noindent Note that all fields are local with respect to the screening currents,
and that all fields are at worst ``half-local'' with respect to any other
field, which makes perfect sense on a Riemann surface which is just a
branched double covering of the complex plane.

For a very thoroughly exposition of the $c=-2$ model see e.g.\
\cite{GFN97}. The field $\mu$ is best viewed as carrying half a branch cut
with it. Therefore, only correlation functions with an even number of these
twist fields (or its excitation $\sigma$) will be non-zero, since these fields
have to come in pairs creating the branch cuts. However, there is no direct
way to tell which twist field joins with which. On the contrary, the
conformal blocks are in one-to-one correspondence with a set of independent
ways of distributing the branch cuts between the twist fields. This is
illuminated by the fact that the operator product expansion of two twist
fields is $[\mu]\star[\mu]=[\mathbb{I}]+[P]$. Indeed, suppose we join two
twist fields with no branch cut between them, which just amounts in connecting
two branch cuts to one, decreasing the genus of the surface by one. The point
in the middle, where the two twist fields joined, can be pulled out of the
new branch cut, and because this point does not have any special property
anymore, it is best described by attaching the identity operator $\mathbb{I}$ 
to it. On the other hand, if two twist fields, which are connected by a branch 
cut, are joined, one branch cut shrinks to zero size, thus changing a genus $g$
surface to a genus $g-1$ surface with one puncture. Clearly, this is the case
where the puncture operator $P$ comes into play. This operator is the
logarithmic partner of the identity and creates the logarithmic divergences
in correlation functions. In fact, we have the following important
2-point functions:
\begin{eqnarray}
  \Vev{\mathbb{I}(z)\mathbb{I}(w)} &=& 0\,,\nonumber\\
  \Vev{\mathbb{I}(z)P(w)}          &=& 1\,,\\
  \Vev{P(z)P(w)}                   &=& -2\log(z-w)\,.\nonumber
\end{eqnarray}

The $c=-2$ theory of the $j=1$ differentials admits a free field 
realization with one scalar field $\varphi_k$ per sheet, $k=0,1$ in the
hyperelliptic case. For later convenience, we now prepare a further
computational tool.
It is possible to use the same scalar fields to construct the $c=1$
conformal field theory of the $j=\frac12$ differentials with fields 
$\Phi_{\tilde{\vec{q}}}(z)$ which carry {\em half\/} the
charges of the $c=-2$ fields $\Phi_{\vec{q}}(z)$ do: If $\vec{q}=(q,0)$,
then $\tilde{\vec{q}}=(\frac{q}{2},-\frac{q}{2})$. To this end, we first
introduce the vertex operators for the branch points for the case that
$j$ is half-integer, which now split into pairs
\begin{eqnarray}
  V_+(a)&=&\exp\left({\rm i}\frac{j}{2}\varphi_0
                    +{\rm i}\frac{j-1}{2}\varphi_1\right)
  \,,\nonumber\\[1mm]
  V_-(a)&=&\exp\left({\rm i}\frac{j-1}{2}\varphi_0
                    +{\rm i}\frac{j}{2}\varphi_1\right)
  \,,
\end{eqnarray}
and define $\Phi_{\tilde{\vec{q}}}=\Phi_{(\pm\frac14,\mp\frac14)}(a)$ as
the corresponding primary fields for the case $j=\frac12$.
In a similar fashion, the other fields $\Phi_{\tilde{\vec{q}}}(z)$ can be 
constructed.\footnote{\,Of course, the charges
of the conjugate fields are different, namely $q^*=-q$, for the $c=1$ theory,
since the background charge $\alpha_0 = 0$ for $c=1$. The $c=1$ CFT with
primary fields given by the $c=-2$ charges can be identified as the CFT
of the Dirac fermion with compactification radius $R=1/\sqrt{2}$. 
Since branch points have now two vertex operators associated
with them, $\Phi_{\frac14}$ and $\Phi_{-\frac14}$, they allow for different
boundary conditions. By saying that we choose the charges to be the same
as for the $c=-2$ theory, we fix the boundary conditions to be the
completely periodic ones.} Hence, we have
$\tilde{\mathbb{I}}=\Phi_{(0,0)}\equiv\mathbb{I}$, 
$\tilde{\mu}=\Phi_{(\frac14,-\frac14)}$,
$\tilde{P}=\Phi_{(\frac12,-\frac12)}$, 
$\tilde{\sigma}=\Phi_{(-\frac14,\frac14)}$, and 
$\tilde{J}=\Phi_{(-\frac12,\frac12)}$, with the conformal weights
$h\in\{0,\frac{1}{16},\frac14,\frac{1}{16},\frac14\}$ respectively.
Clearly, an arbitrary correlation function of the free $c=1$ theory is simply
\begin{equation}\label{eq:free}
  \Vev{\prod_{j=1}^{N}\Phi_{\tilde{\vec{q}}_j}(z_j)}_{c=1} =
  \prod_{i<j}(z_i-z_j)^{\frac12q_iq_j}
\end{equation}
subject to the condition that $\sum_{j=1}^N q_j=0$ (otherwise, appropriate
screening charges must be introduced, or, put in a different language, the
fermionic zero modes must be absorbed by additional fields 
$\phi_k^{(j=\frac12)},\phi_{1-k}^{(1-j=\frac12)}$ inserted into the 
correlator). We may further divide
this $c=1$ theory out of our $c=-2$ theory with the effect that the free
part of correlation functions is canceled,
\begin{equation}\label{eq:vvev}
  \Vvev{\prod_{j=1}^{N}\Phi_{q_j}(z_j)} \equiv \frac{
  \Vev{\prod_{j=1}^{N}\Phi_{(q_j,0)}(z_j)}_{c=-2}}{
  \left(\Vev{\prod_{j=1}^{N}\Phi_{(\frac12q_j,-\frac12q_j)}(z_j)}_{c=1}
  \right)^2}\,.
\end{equation}
These {\em reduced\/} correlators will be the objects we are most interested
in the following. The reason is that we will
express period integrals over meromorphic differential forms in terms of
correlation functions. However, a correlation function always consists
of two parts, firstly the free contribution (sometimes also called the
classical part), and secondly a non-trivial contribution (sometimes called
the quantum part) involving all monodromy properties. This latter part
can be written as certain contour integrals over the screening charges,
if the CFT is a degenerate model, and it is this latter part which will
provide us with the desired information. So, the above construction is
a shorthand for throwing away what we do not need.

The careful reader will note
that naively the charge balance for a $c=-2$ correlator should be
$\sum_{j=1}^N q_j=2\alpha_0=1$ (including all screening charges or zero-mode
absorbing pairs of anti-commuting $j,(1-j)$ fields) 
due to the non-vanishing background charge.
Indeed, as is demonstrated e.g.\ in \cite{GFN97}, correlation functions
of the $c=-2$ theory are non-zero only when one field $P$ is put
at infinity, which nicely accounts for the background charge. The reason is 
that in the $c=-2$ theory we have the non-trivial vacuum structure 
$\langle\mathbb{I}|\mathbb{I}\rangle = 0$, and $\langle P|\mathbb{I}\rangle = 
1$. However, due to the duality between $j$ and $(1-j)$, we could also impose 
the charge balance condition $\sum_{j=1}^N q_j=-2\alpha_0=-1$, which simply
amounts in replacing $\langle -1|$ by $\langle -1|Q_+=\langle 1|$. The
validity of this is ensured, because the screening charge $Q_+$ may act
locally at infinity, and since
$\langle -1|$ is a spin doublet state. This, together
with the fact that, in particular, the branch point field $\mu$ is 
self-conjugate, leads to the following important fact: Any correlation
function $\Vev{\prod_{j=1}^{N}\Phi_{(q_j,0)}(z_j)}_{c=-2}$ contains at least
one conformal block involving precisely
one more screening charge $Q_-$ to ensure the correct charge balance,
than its companion correlation function
$\Vev{\prod_{j=1}^{N}\Phi_{(q_j/2,-q_j/2)}(z_j)}_{c=1}$.
As a consequence, the
above expression (\ref{eq:vvev}) precisely extracts this one integration
$Q_-=\oint J_-(z)$ over the additional screening current.\footnote{\,Note 
that equation (\ref{eq:vvev}) is related to a well known formula
for the determinants of $\bar{\partial}_{(j)}$, namely
$\det_{\vec{m}}\bar{\partial}_{(j=1/2)}\left(\det\bar{\partial}_{(j=0)}
\right)^{1/2}=\theta_{\vec{m}}$, where the characteristics $\vec{m}
=(\vec{m}',\vec{m}'')$ defines the boundary conditions imposed on the
fermions, i.e.\ $j=\frac12$-differentials, and where $\theta_{\vec{m}}$ is
the corresponding theta-constant.
An alternative but equivalent definition can be given, which
entirely remains within the $c=-2$ realm and, moreover, has a direct
physical interpretation in terms of partition functions. Since we want to
perform integrals with just one screening charge, we may write
$\Vvevss{\prod_j\Phi_{q_j}(z_j)} =
\frac{\partial}{\partial\beta}\left.\log\Vev{\prod_j\Phi_{q_j}(z_j)
\,\exp\left(\beta\int J_-(z)\,\hat{c}\right)}\right|_{\beta=0}$. 
Here, $\hat{c}$ stands
for the necessary cocycle which accounts for the shift in momentum. This latter
definition could easily be generalized towards extracting several screening
integrations.}

As a matter of fact, whenever the charge balance of the
primary fields is zero, the denominator of (\ref{eq:vvev}) takes the simple
form (\ref{eq:free}). Otherwise, since we are only interested in such
conformal blocks which involve precisely one screening integration, we will
implicitly assume that the charge balances are ensured via insertion of
appropriate numbers of zero mode absorbing fields, 
$\phi_k^{(j=1)}(Z_m),\phi_{1-k}^{(1-j=0)}(Z'_m)$ in the $j=1$ numerator,
and $\phi_k^{(j=\frac12)}(Z_n),\phi_{1-k}^{(1-j=\frac12)}(Z'_n)$
in the $j=\frac12$ denominator respectively,
and taking a regularized limit $Z_m,Z'_m,Z_n,Z'_n\longrightarrow\infty$.
We remark that the $c=-2$ theory of $j=1$ differentials can formally 
be written as a coset
$(\widehat{SU(2)}_{-1}\times\widehat{SU(2)}_1)/\widehat{SU(2)}_0$, where
the the $\widehat{SU(2)}_1$ stems from the $j=\frac12$ contribution with
$c=1$. It is therefore tempting to interpret equation (\ref{eq:vvev}) 
in the sense that this part can (formally) be factored out. 

Last, but not least, we would like to note how the correlators $\vvev{\cdot}$
behave under a global conformal transformation of the coordinates. Let
$M=\frac{aZ-b}{cZ-d}$ with $ad-bc\neq 0$ be a global conformal transformation
(on the Riemann sphere). A generic CFT correlation function of primary
fields transforms then as
$\Vev{\prod_j\phi_j(z_j)} = \prod_i\left(\left.\frac{\partial M(Z)}{\partial Z}
\right|_{Z=z_i}\right)^{-h_i}\Vev{\prod_j\phi_j(M(z_j))}$. On the other
hand, we have to keep in mind that we throw away the free part of the
correlation functions. The only remaining contributions come from the
integrations with the screening currents. This means that the total
exponents are $-h_{i,(c=-2)} + 2h_{i,(c=1)} = -(q_i^2 - 2\alpha_0q_i)/2 
+ 2q_i^2/4 = q_i/2$. Therefore,
the correlators defined in (\ref{eq:vvev}) transform under $M$ according to
\begin{equation}\label{eq:trafo}
  \Vvev{\prod_{j=1}^{N}\Phi_{q_j}(z_j)} =
  \prod_{i=1}^N\left(\left.\frac{\partial M(Z)}{\partial Z}
  \right|_{Z=z_i}\right)^{\frac{q_i}{2}}
  \Vvev{\prod_{j=1}^{N}\Phi_{q_j}(M(z_j))}\,.  
\end{equation}
If zero-mode absorbing spin $j$ fields, i.e.\ $j$-differentials, are inserted,
these will contribute an additional factor -- in the limit -- 
of $\lim_{z\rightarrow\infty}\left[(\partial_zM(z))\frac{z^2}{M(z)^2}
\right]^{q(j)/2}$ with $z$ the localization variable and
$q(j)$ the charge of the $j$-differential at infinity. For example, to
switch the charge balance condition from $\sum_iq_i=1$ to $\sum_iq_i=-1$,
the zero modes are absorbed with the help of $J_+\sim\ \nop{P^2}$ acting at 
infinity having charge $q(J_+)=2$. We will often omit these zero mode
absorbing fields, since they can be inferred from the charge balance
condition and the requirement that only one screening current integration
is to be performed.

We will later express correlation functions in terms of generalized
hypergeometric functions of several variables. As shown in Appendix B, these
will depend on the {\em inverse\/} crossing ratios, i.e.\ on $M(z_j)^{-1}$. 
Assuming that $M$ maps $\{z_1,z_2,z_3\}\mapsto\{\infty,1,0\}$, the
generalized hypergeometric functions $F$ will depend explicitly only on
$M(z_4)^{-1},\ldots,M(z_N)^{-1}$. Therefore, the above formula applied
to $F$ will be modified to 
\begin{eqnarray}\label{eq:trafo2}
    \Vvev{\prod_{j=1}^{N}\Phi_{q_j}(z_j)} &=&
    \prod_{i=1}^3(\partial_{z_i}M(z_i))^{\frac{q_i}{2}}
    \prod_{i=4}^N\left(\frac{\partial_{z_i}M(z_i)}
    {M(z_i)^{2}}\right)^{\frac{q_i}{2}}
    \lim_{z\rightarrow\infty}\left(z^2\frac{\partial_z M(z)}
    {M(z)^2}\right)^Q\nonumber\\
    &\times& F(M(z_4)^{-1},\ldots,M(z_N)^{-1},M(z\rightarrow\infty)^{-1})\,,
\end{eqnarray}
including the contribution for the zero mode absorbing fields, and
where we have put $Q=1-\frac{1}{2}\sum_iq_i$. We will often
use the notation $x(z)\,\equiv 1/M(z)$ for the inverse crossing ratios.

Similar considerations hold for the power exponents in OPEs, when these are
performed within a $\Vvevss{\cdot}$ correlator. One finds
\begin{eqnarray}\label{eq:vvope}
  \lefteqn{\Vvevss{\ldots\Phi_{q_1}(z)\Phi_{q_2}(w)\ldots}\ \sim
    \phantom{xxxxxxxxxxxxxxxxxxxxxxxxxxxxxx}}
    \nonumber\\
  & &\sum_{{\scriptstyle q\geq 0\atop
            \phantom{\scriptstyle |\{r\}|\geq -1-(\Delta h(q))}}}
    C_{h(q_1),h(q_2)}^{h(q)}(z-w)^{\frac12(q-q_1-q_2)(q-1+q_1+q_2)}
    \nonumber\\[-1.7ex]
  & &\sum_{{\scriptstyle \{r\}\atop
            \scriptstyle |\{r\}|\geq -1-(\Delta h(q))}}
    a_{h(q_1),h(q_2)}^{h(q);\,\{r\}}(z-w)^{1-h(q)_{{\rm min}}}\nonumber\\
  & &\times
    \left(\partial^{-1-(\Delta h(q))-|\{r\}|}
          (z-w)^{h(q)_{{\rm min}}-1}\right)
    \Vvevss{\ldots{\cal L}_{-\{r\}}\Phi_{q}(w)\ldots}\, .
\end{eqnarray}
This form of the OPE also shows the descendant terms and does explicitly 
take into account possible logarithmic divergences due
to {\em different\/} fields of the same conformal scaling dimension (or fields
whose conformal dimensions differ by integers). This happens precisely
when $q$ is a positive integer. We then have that $h(q)=h(q)_{{\rm min}}
+ (\Delta h(q))$ with $h(q)_{{\rm min}}$ the conformal scaling dimension
of the lowest primary field within a Jordan block.
Negative powers of derivatives simply
mean formal term by term integration of a generalized series expansion
(which may contain fractional powers, logarithms, etc.), and
${\cal L}_{-\{r\}}$ is a shorthand for $L_{-r_1^{}}L_{-r_2^{}}\ldots 
L_{-r_n^{}}$. A more rigorous formulation
of the OPE of logarithmic CFTs can be found in \cite{Flo01}.
Linearity is ensured if the OPE is applied before the $\Vvevss{\cdot}$
expectation values are taken.

\subsection{Logarithmic Operators}

The CFT of the spin $(1,0)$ ghost system with central charge $c=-2$ has
the great advantage that the origin of logarithmic operators is well 
understood. It is the presence of the field $\mu$, which in our geometrical
setting simulates branch points, which inevitably leads to the appearance
of the logarithmic operator $P$ and logarithmic divergences in correlation
functions. This holds even in the case of arbitrary $\mathbb{Z}_n$ twists.
Thus, let $\mu_\lambda$ denote the field which simulates a branch point of
ramification number $n$. In this case, the branch point vertex operator
$\Phi_{\vec{q}}$ consists out of $n$ copies of fields $\Phi_{q_k}$ with
the $q_k$ given by Eq.\ \ref{eq:charge}. Looking at just one of these copies,
$\mu_\lambda$ is one of these fields $\Phi_{q_k}$ for a given $k\in\{0,\ldots,
n-1\}$.

We now come to a subtle point concerning the contours, along which the
screening currents are integrated. In our geometrical setting, we would like
to choose contours which are non-trivial elements of the homology. In fact,
it is possible to choose such a contour, if the two fields which constitute
the insertions around which the contour is to be taken, lead to something
local with respect to the screening current. The typical situation in
ordinary conformal field is different, however. To perform a screening
integration, one usually has to take a Pochhammer double loop integral. 
Now, we observe the following: If a twist field $\mu_\lambda$ represents
an, say, $\ell$-fold ramification around a branch point, when this point
is encircled counter-clockwise, then the twist field $\mu_{\lambda^*}$ with
$\lambda^* =  1-\lambda\equiv -\lambda$ mod $1$ represents an $\ell$-fold
ramification backwards, if its insertion point is encircled counter-clockwise.
This is the situation just mentioned, a pair of two twists $\mu_\lambda$ and
$\mu_{\lambda^*}$, which together yield something local with respect to
a simple homology cycle encircling both. A pair of twists $\mu_\lambda$ and
$\mu_{\lambda'}$ with $\lambda'\neq 1-\lambda$ mod $1$ cannot be encircled
by a simple homology cycle, since the single loop cannot close. We depict
the two different situations below, for the sake of simplicity in the
case of $\mathbb{Z}_3$ twists.
$$\includegraphics[width=15cm]{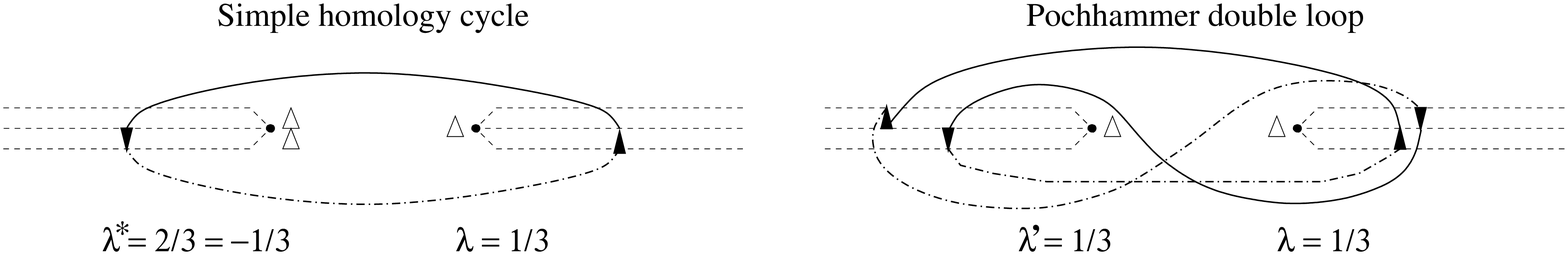}$$
In left situation, we have the possibility to take as screening contour
a simple homology cycle, since the second twist with $\lambda^*=2/3$ undoes the 
effect of the first with $\lambda=1/3$ such that the contour ends on the same 
sheet where it started.
In the right situation, this is not possible, since the second twist cannot
undo the effect of the first, since $\lambda'=\lambda=1/3$ such that
$\lambda'+\lambda\neq 1$. A more careful consideration shows that a single
loop integration is possible whenever $\lambda'+\lambda\in\mathbb{Z}$.

Why does this lead to logarithms? Now, the primary fields $\mu_\lambda$ are
actually superpositions of chiral vertex operators. In the simplest case, the
$\mathbb{Z}_2$ case, the field $\mu_{1/2}$ is actually given by the
superposition $\mu_{1/2}={\cal D}_{1/2,q}^{q+1/2}V_{1/2,q}^{q+1/2}(\cdot,z)
+ {\cal D}_{1/2,q}^{q-1/2}V_{1/2,q}^{q-1/2}(\cdot,z)$ where we have that
$V_{1/2,q}^{q+1/2}(\cdot,z)$ is basically the unscreened vertex operator
$V_{1/2}(z)=\exp({\rm i}\frac{1}{2}\varphi(z))$, while
$V_{1/2,q}^{q-1/2}(\cdot,z)=\oint{\rm d}z'J_(z')V_{1/2}(z)$. This reflects
the fact that the field $\mu_{1/2}$ is degenerate of level two such that
the fusion rules are simply $[\mu_{1/2}]\times[\Phi_q] = [\Phi_{q+1/2}]+
[\Phi_{q-1/2}]$ for any primary field $\Phi_q$. Let us now study the effect
of an operator product expansion of two such fields $\mu_{1/2}$. Geometrically,
this means that we let two branch points run into each other. The unscreened
part of the primary fields $\mu_{1/2}$ will simply add up,
\begin{equation}
  V_{1/2,1/2}^1(|\mu_{1/2}\rangle,z)\ \ :\ \
  \mu_{1/2}(z)\mu_{1/2}(w) \sim (z-w)^{1/4}P(w)\,,
\end{equation}
which geometrically means that the two branch points run into each other to
form a puncture or marked point on the Riemann sphere. The screened part,
however, leads to two different possibilities depending on the choice of the
screening contour. If the screening contour encircles the resulting puncture
at, say, coordinate $z$, all what happens is that we take the residue of
the pole at $z$. The net effect is a constant such that in this situation,
the screened part results simply in $Q_-P(z)=\oint z^{-1}\mathbb{I}=
\mathbb{I}$, the identity
operator. On the other hand, if the contour is chosen such that it
gets pinched between the two branch points, a single cut survives. 
Let us assume without loss of generality that the puncture is created by
letting a branch point at $z$ run into a branch point at zero. The
pinching of the contour leads to an divergence for $z\rightarrow 0$ which
reflects in its behavior the appearance of the cut. The integration around
the pole cannot be performed to simply yield a residue. Instead, the
defect of doing so depends on how close the two branch points get when
pinching the contour. Since the contour cannot close, the result must
be given by the indefinite integral of $1/u$, which is $\int^{(z-w)} u^{-1}=
\log(z-w)$. 
$$\includegraphics[width=15cm]{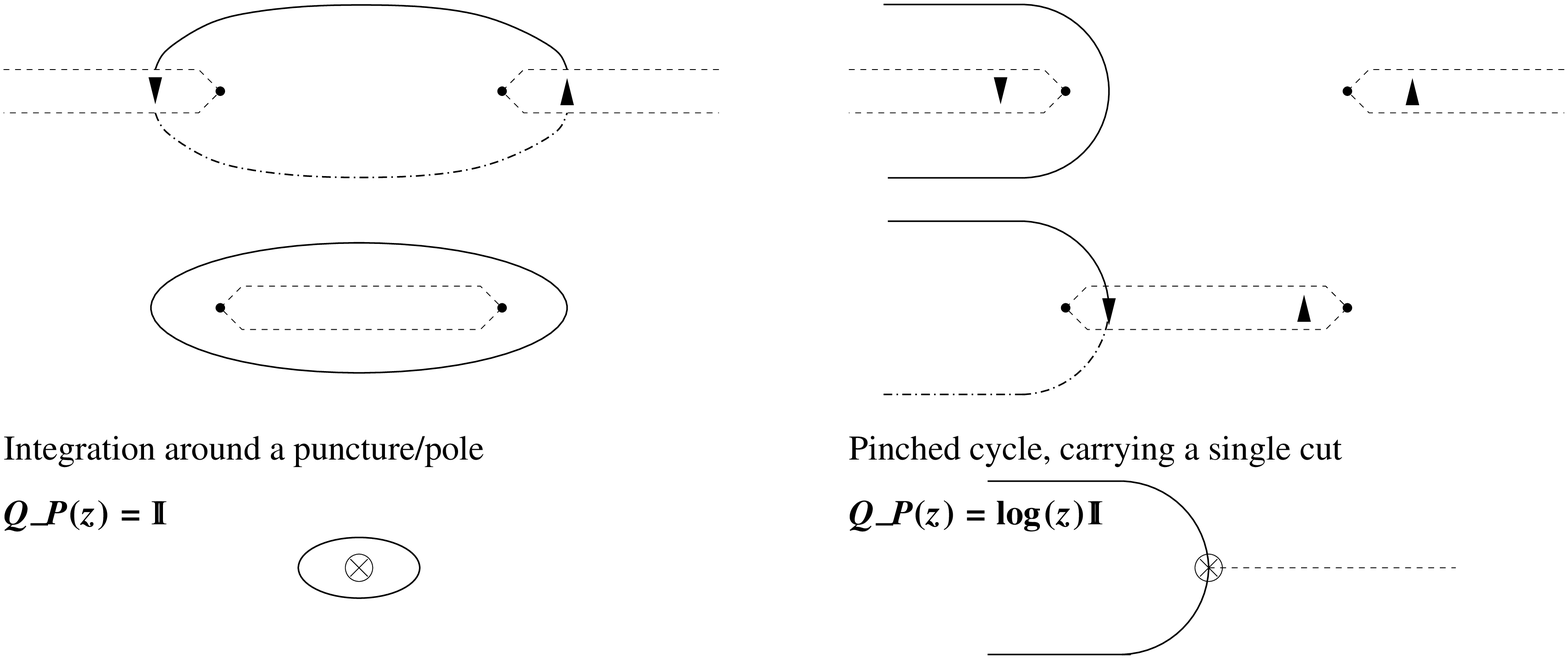}$$
A similar argument holds for arbitrary twists $\mu_\lambda$ in the case
where the pinching is done with $\mu_{\lambda^*}$ resulting in the
indefinite integral of $z^{-\lambda-\lambda^*}=z^{-1}$, while a pinching
of a contour between $\mu_\lambda$ and $\mu_{\lambda'}$ for $\lambda'\neq
\lambda^*$ does not lead to a logarithm since the indefinite integral now is
$\int z^{-\lambda-\lambda'}=-\frac{1}{\lambda+\lambda'-1}
z^{1-\lambda-\lambda'}$. 
Furthermore, the primary field $P$ is again a
superposition of chiral vertex operators. Since it is degenerate of level
three, the superposition contains three terms with zero, one or two
screening integrations attached, respectively. Again, integer order poles
may arise leading to logarithms, if one contour is pinched between two
of the puncture operators.
Indeed, the integral one has actually to perform can be brought into the form
\begin{equation}
  \int^{2\epsilon}{\rm d}u\frac{1}{(u-\epsilon)^\lambda(u+\epsilon)^{\lambda'}}
  \sim\int^{2\epsilon}{\rm d}u\left(\frac{1}{u^{\lambda+\lambda'}} +
  \frac{\lambda-\lambda'}{u^{\lambda+\lambda'+1}}\epsilon+{\cal O}(
  \frac{\epsilon^2}{u^{\lambda+\lambda'+2}})\right)\,,
\end{equation}
which develops a simple pole precisely for $\lambda'=\lambda^*=1-\lambda$,
yielding then a logarithmic divergence $\log(2\epsilon)$. Note that, would
we have used a Pochhammer double loop, the logarithmic divergence would appear
twice, but with opposite signs due to the fact that each loop comes with both
orientations. It is therefore crucial that we use homology cycles as the
contours for screening current integrations.

As a demonstration, 
let us compute the possible conformal blocks of the four-point functions
$\langle\mu_{1/2}(\infty)\mu_{1/2}(1)\mu_{1/2}(x)\mu_{1/2}(0)\rangle$.
First of all, we have basically two inequivalent ways to join the two pairs
of branch points by cuts. Secondly, the homology for the torus is spanned by
two elements, $\alpha$ and $\beta$. We have depicted canonical choices for 
the homology cycles $\alpha$ and $\beta$ for both configurations for the 
branch cuts.
$$\includegraphics[width=15cm]{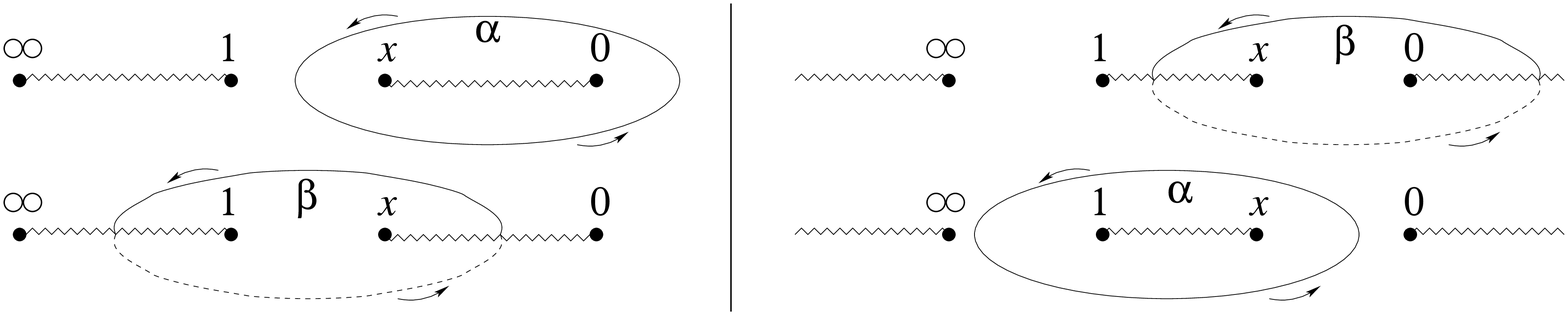}$$
We now insert the OPE for $|x|\ll1$ such that essentially try to reduce the
four-point functions to a three-point function. In the left configuration,
the screened vertex operator at $x$ leads after insertion of the OPE to
\begin{equation}
  V_{1/2,1/2}^0(|\mu_{1/2}\rangle,x)\ \ :\ \
  \mu_{1/2}(x)\mu_{1/2}(0) \sim x^{1/4}\oint_\alpha{\rm d}u u^{-1}\mathbb{I}
  \sim x^{1/4}\mathbb{I}
\end{equation}
for the homology cycle $\alpha$, and for the $\beta$ cycle, the pinching leads
to the result
\begin{eqnarray}
  V_{1/2,1/2}^0(|\mu_{1/2}\rangle,x)\ \ :\ \
  \mu_{1/2}(x)\mu_{1/2}(0) &\sim& x^{1/4}\oint_\beta{\rm d}u u^{-1}\mu_{1/2}(x)
  \mu_{1/2}(0)\nonumber\\
  &\sim& x^{1/4}\int^x\!\!{\rm d}u u^{-1}\mathbb{I} = x^{1/4}\log(x)\mathbb{I}
  \,.\phantom{xxxx}
\end{eqnarray}
Note that there can be only an even number of pinchings. Thus, 
logarithms as the one in the last OPE will show up in correlation functions
always as pairs. Since the logarithmic divergence is also dependent on
the orientation of the contour, any contour pinched twice will lead to
the difference of two logarithms.

Therefore, contracting further for the left configuration and the
$\alpha$ cycle leads to
\begin{eqnarray}
  \langle\mu(\infty)\mu(1)\mu(x)\mu(0)\rangle_\alpha &\sim&
  x^{1/4}\langle\mu(\infty)\mu(1)(Q_-P(0)\rangle
  \sim x^{1/4}\langle\mu(\infty)\mu(1)\mathbb{I}\rangle
  \nonumber\\[1mm]
  &\sim& x^{1/4}\langle \mu(\infty)\mu(0)\rangle
  \sim x^{1/4}\langle P(0)\rangle = x^{1/4}
\end{eqnarray}
to leading order, where no logarithms show up, since the contour integration
can be performed in a region where no pinching occurs. 
All operators here are to be understood as unscreened
vertex operators, since the screening charge has been explicitly inserted.
Since there is only one screening allowed due to charge balance, the OPE
inserted in the last line yields only one term.
Note that $\langle\mathbb{I}\rangle=0$ and
$\langle P\rangle = 1$. In case of the $\beta$ cycle, we obtain to leading
order
\begin{eqnarray}
  \langle\mu(\infty)\mu(1)\mu(x)\mu(0)\rangle_\beta &\sim&
  x^{1/4}\langle\mu(\infty)[\mu(1)Q_-|_xP(0)+Q_-|_1P(x)\mu(0)]\rangle
  \nonumber\\[1mm]  &\sim&
  x^{1/4}\langle\mu(\infty)[\mu(1)\log(x)\mathbb{I}(0)
  -\log(1-x)\mathbb{I}(x)\mu(0)]\rangle\nonumber\\[1mm] & \sim&
  x^{1/4}(\log(x)-\log(1-x))\langle\mu(\infty)\mu(0)\rangle\\ & \sim&
  x^{1/4}(\log(x)-\log(1-x)\langle P(0)\rangle = x^{1/4}\log\Big(\frac{x}{1-x}\Big)\,.
  \nonumber
\end{eqnarray}
Here, we have tried to indicate, where the screening contour gets pinched.
Note that in case of the $\beta$ cycle, two pinchings occur and a relative
sign must be taken into account. To leading
order, this yields the well-known results $x^{1/4}\log(x)$ for $|x|\ll 1$.
Of course, we have made use of translational invariance of the correlators
to arrive at the second-last line.

The right configuration can be inferred in a completely analogous way.
However, it is equivalent to the left configuration if the region
$|1-x|\ll 1$ is considered instead. This essentially exchanges the role
of the homology cycles $\alpha$ and $\beta$. We thus see that there are
two inequivalent conformal blocks in one-to-one correspondence to the
basis of homology cycles. The choice of the screening contour implies
the choice of the correct screened vertex operators which build up the
primary fields, i.e.\ the choice of internal channel in the OPE.

As a last remark, we note that the logarithmic divergence also causes the
resulting representation $[P]$ to be indecomposable. This can be seen from
the fact that $L_0$ is the generator of both dilatations, and rotations. The
latter, in turn, generate monodromy transformations. Since $\log(x)$ is
multivalued, we see that for a pinched contour, we must have
$L_0Q_-P(x) = Q_-P(x)+\mathbb{I}$ which is precisely the statement that
$L_0|P\rangle = |0\rangle$ forms a Jordan cell.

\subsection{Holomorphic Differentials}

Now, we have all the important data to start expressing differential forms on 
hyperelliptic curves by correlation functions of the $c=-2$ theory. To
demonstrate this, we start with a very simple example, namely the canonical
holomorphic differentials $\omega_n=z^n\,y^{-1}\,{\rm d}z$ with 
$n=0,\ldots,g-1$,
and $y$ given by (\ref{eq:curve}). These forms are also called Abelian
forms of the first kind. We claim that the holomorphic 
differentials are simply represented by 
\begin{equation}
  \omega_n = \nop{J(0)^n}\,\prod_{k=1}^{2g+2}\mu(e_k)\,J_-(z)
           = \Phi_{-n}(0)\prod_{k=1}^{2g+2}\Phi_{1/2}(e_k)\,J_-(z)\,,
\end{equation}
up to a contribution $\nop{(\phi^{(j=0)}(\infty))^{n+1-g}}$ of zero mode
absorbing fields. 

By this, we mean the following: First, the normal ordered product of $n$
currents projects down to the primary field of charge $q=n$ and conformal
weight $h_{n+1,1}=n(n+1)/2$. Next, integrating $\omega_n$ along a cycle of 
the hyperelliptic curve simply results in integrating the screening current
along this cycle, i.e.
\begin{equation}
  \oint_{\alpha}\omega_n = 
  \left.\Phi_{-n}(0)\prod_{k=1}^{2g+2}\Phi_{1/2}(e_k)\,Q_-\right|_{\alpha}
  \,,
\end{equation}
where $Q_-$ denotes the screening charge operator $Q_-=\oint J_-(z)$.
This results just in the non-free part of the correlation function
$\Vev{\Phi_{-n}(0)\prod_{k=1}^{2g+2}\Phi_{1/2}(e_k)}_{c=-2}$, actually in a
certain linear combination with integer coefficients of a basis of conformal
blocks of this correlation function, depending on the cycle $\alpha$.
Choosing $\alpha$ as an element of a basis of cycles, and keeping in mind
equation (\ref{eq:vvev}), we finally find
\begin{equation}\label{eq:periods}
  \oint_{\alpha}\omega_n = 
    \Vvev{\nop{P^{n+1-g}(\infty)}\;
      \Phi_{-n}(0)\prod_{k=1}^{2g+2}\Phi_{1/2}(e_k)}_{(\alpha)}\,,
\end{equation}
where we {\em define\/} the basis of conformal blocks, labeled by $\alpha$,
to be in correspondence with the chosen basis of cycles. Of course, 
$\nop{P^m(\infty)}$ absorbs any zero modes and could equally been written
as $\nop{(\phi^{(j=0)}(\infty))^m}$, and it is always implicitly understood
that the projection of the normal ordered product onto the primary field
$\Phi_{m}(\infty)$ is taken.

It is now worth noting that the primary field 
$\mu=\Phi_{1/2}\equiv\Psi_{1,2}$, which
creates (half of) the branch cuts, actually is a degenerate field of
level 2. It follows that the elements of the period matrix satisfy certain
partial differential equations of second order (as well as equations of
first order induced by the conformal Ward identities). Reexpressing these
equations in the moduli of the hyperelliptic curve (instead of its branch
points) relates them to the Picard-Fuchs equations for the periods.

Clearly, more general forms can be considered, and the most general
Abelian differential can be brought into the form (up to zero mode
absorbing fields)
\begin{eqnarray}
    \omega &=& \frac{\prod_{i=1}^{M}(z-z_i)}{
    \prod_{k=1}^{2g+2}\sqrt{z-e_k}\prod_{j=1}^{N}(z-p_j)}\,{\rm d}z\nonumber\\[1mm] 
    &=& 
    \prod_{i=1}^{M}\Phi_{-1}(z_i)\prod_{k=1}^{2g+2}\Phi_{1/2}(e_k)
    \prod_{j=1}^{N}\Phi_{1}(p_j)\,J_-(z)\,,
\end{eqnarray}
where the $z_i$ and $p_j$ do not need to be all disjunct. In this latter
case, it is understood that products of fields at the same point are
normal ordered. Also, we do not exclude the case that not all the $z_i$
are different from the $e_k$. Since $P=
\Phi_{1}\equiv\Psi_{1,3}$ is a degenerate primary field of level 3, we 
immediately see that
in the case of meromorphic differentials with poles not at infinity, the
periods of such a differential also must satisfy partial differential 
equations of third order. This is reflected in the fact that the Picard-Fuchs
equations for Seiberg-Witten models with massive hypermultiplets are of
third order. However, as long as there is only one pole,
the periods are still uniquely determined by second order equations. Such
forms are related to Abelian forms of the second kind. In fact, since 
all $n$-point functions of a two-dimensional conformal field theory are
in principle determined by its 2-, 3- and 4-point functions, an Abelian
differential of the third kind, written as an $n$-point function, can
be expressed solely with 4-point functions having at least one field
degenerate of level 2 in them.

To effectively compute a general Abelian form which also has zeroes,
it is convenient to adopt the usual conformal field theory point of view:
Writing the form as
\begin{equation}
  \omega = \prod_{i=1}^{K}(z-a_i)^{r_i}\,,
\end{equation}
the $r_i$ would be restricted to be in $\{-1/2,+1/2,-1,+1\}$. For the
correlation functions, it is better to keep the $r_i$ as variable,
compute the screening charge integral in a region of the set of exponents
$r_i$ where everything converges, and then perform an analytic continuation
to the correct values of the $r_i$. This procedure is particularly helpful
for the case that $\omega$ has zeroes. Instead of performing an integration
along a path that joins two of these zeroes, we might introduce new cycles
which go around two of the zeroes, and perform the calculation as if all
the $r_i$ were negative. This usually leads to generalized hypergeometric
or hyperelliptic integrals within a region of parameter space there the
integrals are well-defined and converge. Other regions of the parameter space
$\{r_i\}$ can be reached by analytic continuation.

\section{Seiberg-Witten Solutions of Supersymmetric Four-Dimen\-sional 
         Yang-Mills Theories}

In a much celebrated work \cite{SW94}, 
Seiberg and Witten found an exact solution
to $N$=$2$ supersymmetric four-dimensional Yang-Mill theory with gauge group
$SU(2)$. This paper initiated a whole new, tree sized, branch of research
leading to a vast set of exactly solvable Yang-Mills theories
in various dimensions and with various degrees of supersymmetry. For some
basic or introductory works see, for example, \cite{SW1,SW2,SW3,SW4,SW5,SW6} and
references therein. Of
particular interest for these solutions is the understanding of the moduli
space of vacua, which in many cases turns out to be a hyperelliptic
Riemann surface. In particular, simply-laced Lie groups lead to spectral
curves which are hyperelliptic.

The BPS spectrum of such a model is entirely determined by the periods of
a special meromorphic 1-differential on this Riemann surface, the famous
Seiberg-Witten differential $\lambda_{{\rm SW}}$. A general hyperelliptic
Riemann surface can be described in terms of two variables $w,Z$ in the
polynomial form 
\begin{equation}\label{eq:surf}
    w^2 + 2A(Z)w + B(Z) = 0
\end{equation}
with $A(Z),B(Z)\in\mathbb{C}[Z]$. After
a simple coordinate transformation in $y=w-A(Z)$, this takes on the 
more familiar form $y^2 = A(Z)^2 - B(Z)$. But we might also write
the hyperelliptic curve in terms of a rational map if we divide the
defining equation (\ref{eq:surf}) by $A(Z)^2$ and put $\tilde{w}=w/A(Z)+1$
to arrive at the representation
\begin{equation}\label{eq:ratmap}
    (1-\tilde{w})(1+\tilde{w}) = \frac{B(Z)}{A(Z)^2}\,.
\end{equation}
This form is very appropriate in the frame of Seiberg-Witten models, since
the Seiberg-Witten differential can be read off directly: The rational map
$R(Z)=B(Z)/A(Z)^2$ is singular at the zeroes of $B(Z)$ and $A(Z)$, and is
degenerate whenever its Wronskian $W(R) \equiv W(A(Z)^2,B(Z)) =
(\partial_ZA(Z)^2)B(Z) - A(Z)^2(\partial_ZB(Z))$ vanishes. This is precisely
the information encoded in $\lambda_{{\rm SW}}$ which for arbitrary
hyperelliptic curves, given by a rational map $R(Z)=B(Z)/A(Z)^2$, can be
expressed as
\begin{equation}\label{eq:lamSW}
    \lambda_{{\rm SW}} = 
    \frac{Z}{2\pi{\rm i}}\,{\rm d}(\log\frac{1-\tilde{w}}{1+\tilde{w}}) =
    \frac{1}{2\pi{\rm i}}{\rm d}(\log R(Z))\frac{Z}{\tilde{w}}
    = \frac{1}{2\pi{\rm i}}
    \frac{W(A(Z)^2,B(Z))}{A(Z)B(Z)}\frac{Z\,{\rm d}Z}{y}\,.
\end{equation}
Note that the fact that the denominator polynomial is a square guarantees
the curve to be hyperelliptic. It is this local form of the Seiberg-Witten
differential which serves as a metric ${\rm d}s^2 = |\lambda_{{\rm SW}}|^2$ 
on the Riemann surface. And it is this local form which arises as the 
tension of self-dual strings coming from 3-branes in type II string theory
compactifications on Calabi-Yau threefolds.\footnote[1]{\,This derivation 
of the Seiberg-Witten differential is equivalent 
to the one from integrable Toda systems with spectral curve $z+1/z+r(t) = 
z+1/z+2A(t)/\sqrt{B(t)}=0$, where $\lambda_{{\rm SW}}=t\,{\rm d}(\log\,z)$ 
is nothing else than the Hamilton-Jacobi function of the underlying integrable 
hierarchy. However, the price paid for this very simple form of $\lambda_{{\rm
SW}}$ is that $r(t)$ is now only a fractional rational map.}

Let us, for the sake
of simplicity, concentrate on $N$=$2$ $SU(N_c)$ Yang-Mills theory with
$N_f$ massive hypermultiplets. Then, the hyperelliptic curve $y^2=A(x)^2-B(x)$
takes the form
\begin{equation}
  y^2 = \left(x^{N_c} - \sum_{k=2}^{N_c}s_k x^{N_c-k}\right)^2 
          -\Lambda^{2N_c-N_f}\prod_{i=1}^{N_f}(x-m_i)
        = \prod_{j=1}^{2N_c}(x-e_j)\,,
\end{equation}
where we have absorbed any dependency of $A(x)=\prod_{k=1}^{N_c}(x-a_k)$ on 
the $m_i$, which is the case for $N_f > N_c$, in a redefinition of the $a_k$ 
or $s_k$ respectively. Then, the Seiberg-Witten differential takes the
general form
\begin{equation}
   \lambda_{{\rm SW}}(SU(N_c)) = \frac{1}{2\pi{\rm i}}
    \frac{x\prod_{l=1}^{N_c+N_f-1}(x-z_l)}{
    \prod_{j=1}^{2N_c}\sqrt{x-e_j}\prod_{i=1}^{N_f}(x-m_i)}\,{\rm d}x\,,
\end{equation}
where the $z_l$ denote the zeroes of $2A(x)'B(x)-A(x)B(x)'$.
As a result, the total order of the general Seiberg-Witten form 
(\ref{eq:lamSW}) vanishes,
$(1 + N_c+N_f-1)\cdot(1) + (2N_c)\cdot(-\frac12) + (N_f)\cdot(-1) = 0$,
meaning that the charge balance for the corresponding primary fields is 
identical zero. Hence, the denominator in (\ref{eq:vvev}) is just the
free part (\ref{eq:free}).
The periods of this differential along a basis of cycles which can
be chosen in such a way that they encircle pairs $(e_i,e_j)$ are then given
by the conformal blocks (including a factor $J_+(\infty)\ \sim\ 
\nop{P(\infty)^2}\ \sim\ \Phi_{2}(\infty)$ for the double pole of
the Seiberg-Witten differential at infinity)
\begin{eqnarray}
   a_{(e_i,e_j)} &=& \oint_{C_{(e_i,e_j)}}\lambda_{{\rm SW}}(SU(N_c))
   \nonumber\\ &=& 
    \frac{1}{2\pi{\rm i}}\Vvev{\prod_{j=1}^{2N_c}\mu(e_j)\prod_{i=1}^{N_f}P(m_i)
    \prod_{l=0}^{N_c+N_f-1}J(z_l)\;\nop{P(\infty)^2}}_{(e_i,e_j)}\!\! ,
\end{eqnarray}
where we have defined $z_0=0$. However, once we live in the CFT picture,
we may also define ``periods'' which run along cycles encircling two
zeroes of $\lambda_{{\rm SW}}$. We will investigate this in more detail
with the help of an explicit example.

But before we do so, we would like to make some more general remarks.
First of all, the K\"ahler potential for the metric on the field space of
$N\!=\!2$ supersymmetric Yang-Mills theories, $K(a_i,\bar a_j) =
\frac{{\rm i}}{2}((\partial_{\bar a_j}\bar{{\cal F}})a_i 
- (\partial_{a_i}{\cal F})\bar a_j)$ with ${\cal F}(\{a_i\})$ the 
holomorphic prepotential, precisely resembles the single valued
conformally invariant combinations of left and right chiral conformal
blocks,
\begin{equation}
  K_i^{\phantom{i}j} = \frac{{\rm i}}{2}(a_i\bar a_D^j - \bar a_i a_D^j)\,.
\end{equation}
This unusual off-diagonal combination of conformal blocks is a typical
feature of logarithmic CFTs. The full single-valued correlation
function is nothing else than $\vvev{|\lambda_{{\rm SW}}|^2}={\rm tr}\,K$.
Since the periods of the holomorphic 1-forms are also expressed in terms
of conformal blocks, similar results hold for the metric on field space
itself, 
\begin{equation}
  ({\rm d}s)^2 = \IM\,{\rm d}a_D^i\,{\rm d}\bar a_i
\end{equation}
when the exterior derivatives are expressed as derivatives with respect
to the moduli $u_k$ or $s_k$. Therefore, the meromorphic 1-form
$\lambda_{\rm SW}$ is a generating differential, and the $SL(2,\mathbb{C})$
invariance of the metric is nothing else than the conformal invariance of 
the correlation functions of our CFT. Note that again $({\rm d}s)^2$ is
identical to the full single-valued correlation function combined out
of the conformal blocks. More generally, we can define for an arbitrary
Abelian differential $\Omega$ the analogue of the K\"ahler potential,
${\cal K}(\Omega)_i^{\phantom{i}j} = \frac{i}{2}(\oint_{\alpha_i}\Omega
\oint_{\beta^j}\bar{\Omega} - \oint_{\alpha_i}\bar{\Omega}\oint_{\beta^j}
\Omega)$. Moreover, duality in $N\!=\!2$
supersymmetric Yang-Mills theories is nothing else than crossing symmetry
of 4-point (and higher-point) functions. 
Finally, the coupling constants are given in terms of the period matrix.
On a genus $g$ Riemann surface, the period matrix is defined for a canonical
symplectic basis of cycles $\{\alpha_i,\beta^i\}_{1\leq i\leq g}$ and a
basis of the holomorphic 1-differentials $\{\omega_i\}_{1\leq i\leq g}$ by the
$(g,2g)$ matrix
\begin{equation}
  (\Pi_{D\,j}^{\phantom{D\,j}i},\Pi^{}_{ij}) = 
  \left(\oint_{\beta^i}\omega_j,\oint_{\alpha_i}\omega_j\right)\,.
\end{equation}
So, the coupling constants are simply $\tau^{ik} = (\Pi^{-1}\Pi_D)^{ik}$,
and thus expressed in terms of correlation functions.
For completeness, we also mention that the first and second Riemann
bilinear relation, $\tau-(\tau)^t = 0$ and $\IM(\tau)>0$, are immediate
consequences.

\subsection{Periods of the Seiberg-Witten Differential}

We will look at the simplest example first, the Seiberg-Witten model
with gauge group $SU(2)$. The resulting Riemann surface is then simply of
genus one, a torus.

Let us start with a warm up by calculating the periods of the
only holomorphic one-form for the torus.
The torus in question is given by $y^2 = (x^2-u)^2 - \Lambda^4$
with the four branch points $e_1=\sqrt{u-\Lambda^2}$,
$e_2=-\sqrt{u+\Lambda^2}$, $e_3=-\sqrt{u-\Lambda^2}$, $e_4=\sqrt{u+\Lambda^2}$.
The standard periods of the holomorphic form ${\rm d}x/y$
are easily computed (where the normalization has been fixed to be in
accordance with the asymptotic behavior of $a$ and $a_D$ in the weak
coupling region).
With $\xi=1/M(e_4)=\frac{(e_1-e_4)(e_3-e_2)}{(e_2-e_1)(e_4-e_3)}$ the
inverse crossing ratio,
$\xi=(u-\sqrt{u^2-\Lambda^4})/(u+\sqrt{u^2-\Lambda^4})$, we have
\begin{eqnarray}
   \pi_1 = \frac{\partial a}{\partial u}
        &=& \frac{\sqrt{2}}{2\pi}\int_{e_2}^{e_3}\frac{{\rm d}x}{y}
       \ =\ \frac{\sqrt{2}}{2\pi}
            \Vvevss{\mu(e_1)\mu(e_2)\mu(e_3)\mu(e_4)}_{(e_2,e_3)}
            \nonumber\\[1mm]
        &=& \frac{\sqrt{2}}{2\pi}
            (e_3-e_2)^{-\frac{1}{2}}(e_4-e_1)^{-\frac{1}{2}}
            \Vvevss{\mu(\infty)\mu(1)\mu(0)\mu(M(e_4))}_{(0,1)}
            \nonumber\\[1mm]
        &=& \frac{\sqrt{2}}{2}
            (e_2-e_1)^{-\frac{1}{2}}(e_4-e_3)^{-\frac{1}{2}}
            {}_2F_1({\textstyle\frac12,\frac12};1;\xi)\,.
\end{eqnarray}
The other period is obtained in complete analogy by exchanging $e_2$ with
$e_1$, yielding
\begin{equation}
  \pi_2 = \frac{\partial a_D}{\partial u} =
        \frac{\sqrt{2}}{2\pi}\int_{e_1}^{e_3}\frac{{\rm d}x}{y} =
        \frac{\sqrt{2}}{2}(e_1-e_2)^{-\frac{1}{2}}(e_4-e_3)^{-\frac{1}{2}}
        {}_2F_1({\textstyle\frac12,\frac12};1;1-\xi)\,.
\end{equation}
Here and in the following, (generalized) hypergeometric functions with
arguments such as $1-\xi$ are understood as expansions around $1-\xi$ and
should be analytically continued to a region around $\xi$. This will result in
the desired logarithmic divergences. For example, with the usual
Frobenius process we find (the factor $\pi=\Gamma(\frac12)^2$ stems from
the formula for analytic continuation of hypergeometric functions)
\begin{eqnarray}
  \pi\,{}_2F_1({\textstyle\frac12,\frac12;1;1\!-\!\xi}) &=&
    {}_2F_1({\textstyle\frac12,\frac12;1;\xi})\log(\xi) +
    \sum_{n=0}^{\infty}\left.\left(\frac{\partial}{\partial\varepsilon}
    \frac{(\frac12\!+\!\varepsilon)_n(\frac12\!+\!\varepsilon)_n}
    {(1\!+\!\varepsilon)_n(1\!+\!\varepsilon)_n}
\right)\right|_{\varepsilon=0}\!\!\xi^n
    \nonumber\\[1mm]
    &=& {}_2F_1({\textstyle\frac12,\frac12;1;\xi})\log(\xi)\nonumber\\[2mm] 
    & &{ }+\left.\partial_{\varepsilon}\,
    {}_3F_2({\textstyle 1,\frac12+\varepsilon,\frac12+\varepsilon;
    1+\varepsilon,1+\varepsilon;\xi})\right|_{\varepsilon=0}\,.
\end{eqnarray}
These results are, of course, well known. Less known might be the
fact that for the case without hyper-multiplets, $N_f=0$, we can express
the periods of the Seiberg-Witten form by the Lauricella function
$F_D^{(3)}$. In fact,
\begin{eqnarray}
      a(u) &=& \frac{\sqrt{2}}{2\,\pi}\int_{e_2}^{e_3}\frac{4x^2\,{\rm d}x}{y}
         \ =\ \frac{2\sqrt{2}}{\pi}
             \Vvevss{\Phi_2(\infty)
               \mu(e_1)\mu(e_2)\mu(e_3)\mu(e_4)\Phi_{-2}(0)}_{(e_2,e_3)}
             \nonumber\\[1mm]
         &=& \frac{2\sqrt{2}}{\pi}\,\frac{e_1^2}
             {(e_3-e_2)^{\frac12}(e_4-e_1)^{\frac12}}\nonumber\\[2mm]
         & & \times\Vvevss{\mu(\infty)\mu(1)\mu(0)\mu(M(e_4))\Phi_{-2}(M(0))
               \Phi_2(M(\infty))}_{(0,1)}
             \nonumber\\[1mm]
         &=& 2\sqrt{2}\frac{e_3^2}
             {(e_4-e_3)^{\frac12}(e_2-e_1)^{\frac12}}
             F_D^{(3)}({\textstyle\frac12,\frac12},-2,2,1;\xi,\eta,\varpi)\,,
\end{eqnarray}
with the second inverse cross ratio $\eta=1/M(0)=
\frac{e_1(e_2-e_3)}{(e_1-e_2)e_3}$, and $\varpi=1/M(\infty)=
\frac{e_2-e_3}{e_2-e_1}$ the
inverse of the image of the double pole at infinity
(which absorbs the zero modes). 
The Lauricella $D$-type functions are generalized hypergeometric
functions in several variables. We collect all we need about them in 
Appendix~A. 
For $n=1$, they reduce to the ordinary Gauss hypergeometric
functions ${}_2F_1(a,b_1;c;x_1)$, and for $n=2$, they are nothing else than
the Appell functions $F_1(a;b_1,b_2;c;x_1,x_2)$. A great deal of information
on these functions may be found for example in the book \cite{Ext72} by
Exton. An important fact is that $F_D^{(n)}$ satisfies the following system
of partial differential equations of second order:
\begin{eqnarray}
    {\cal D}_j &=&
    (1\!-\!x_j)\sum_{k=1}^nx_k\,\frac{\partial^2}{\partial x_k\partial x_j}
    + \left(c\!-\!(a\!+\!b_j\!+\!1)\,x_j\right)\frac{\partial}{\partial x_j}
    - b_j \sum_{{\scriptstyle k=1\atop\scriptstyle k\neq j}}^n
      x_k\frac{\partial}{\partial x_k}
    \!-\! ab_j\,,\nonumber\\
    0 &=& {\cal D}_j\,F_D^{(n)}\,,
\end{eqnarray}
where $j=1,\ldots,n$. Interestingly, this remains true even in the case
that massive hypermultiplets are present ($N_f>0$), while the Picard-Fuchs
equations now are of third order. However, the price paid is an artificially
enlarged number of variables. Furthermore, we easily can write down
differential equations of second and third order for each field in
the correlator which is proportional to $F_D^{(n)}$, depending on whether
the field is degenerate of level two, e.g.\ $\mu=\Psi_{1,2}$,
$\Phi_{-1}=\Psi_{2,1}$, or three as $\Phi_1=\Psi_{1,3}$ (where we consider the
$c=-2$ CFT as the degenerate model with $c=c_{2,1}$) according to \cite{BPZ83}.

Again, we may obtain the dual period by exchanging $e_2$ with $e_1$,
yielding
\begin{equation}
      a_D(u) = 2\sqrt{2}\,\frac{e_3^2}
             {(e_4-e_3)^{\frac12}(e_1-e_2)^{\frac12}}\,
             F_D^{(3)}({\textstyle\frac12,\frac12},-2,2,1;1-\xi,1-\eta,
             1-\varpi)\,.
\end{equation}
The two periods given above are by construction the
$a_{(\alpha)}$ and $a_{(\beta)}$ periods respectively. 
In the same way, we
find the period integrated between $e_2$ and $e_4$, which is
\begin{equation}\label{eq:adyon}
    a_{(2\alpha-\beta)}(u) = 2\sqrt{2}\frac{-e_2^2}
              {(e_4-e_3)^{\frac12}(e_1-e_2)^{\frac12}}
              F_D^{(3)}({\textstyle\frac12,\frac12,-2,2,1;
              1-\xi,\frac{\xi-1}{\eta-1},\frac{\xi-1}{\varpi-1}})
              \,.
\end{equation}
It is worth
noting that the dependency on three variables is superficial, since all
cross ratios are solely functions in the four branch points. Indeed,
we have $\xi=\varpi^2$, $\eta=-\varpi$. 
The inverse crossing ratios have the nice property that they
tend to zero for $|u|\!\gg\!1$, e.g.\ $\xi\sim (\frac12\frac{\Lambda^2}{u})^2
+ O(u^{-4})$. Hence,
the overall asymptotics of $a(u)$ and $a_D(u)$ is entirely determined by
the prefactors, which are $a(u)\sim
\frac{2\sqrt{2}e_3^2}{\sqrt{e_4-e_3}\sqrt{e_2-e_1}}
\sim \sqrt{2u} + O(u^{-\frac12})$ and $a_D(u)\sim
\frac{\sqrt{2}e_3^2}{\pi\sqrt{e_4-e_3}\sqrt{e_1-e_2}}\log(\xi)\sim
\frac{{\rm i}}{\pi}\sqrt{2u}\log(u) + O(u^{-\frac12}\log(u))$. Expanding
$a(u)$ as a power series in $1/u$ yields the familiar result
\begin{eqnarray}
      a(u) &=& \sqrt{2u}\left[1-\frac{1}{16}\frac{\Lambda^4}{u^2}
             -\frac{15}{1024}\frac{\Lambda^8}{u^4}
             -\frac{105}{16384}\frac{\Lambda^{12}}{u^6}
             -\frac{15015}{4194304}\frac{\Lambda^{16}}{u^8} + O(u^{-10})
             \right]\nonumber\\[1mm]
         &=& \sqrt{2}\sqrt{u+\Lambda^2}\,{}_2F_1({\textstyle-\frac12,\frac12,1};
             \frac{2\Lambda^2}{u+\Lambda^2})\,.
\end{eqnarray}

\subsection{Asymptotics and OPEs}

The strength of the CFT picture becomes apparent when asymptotic regions
of the moduli space are to be explored. Then, OPE and fusion rules provide
easy and suggestive tools. For example, the asymptotics of $a(u)$ and
$a_D(u)$ follow directly from the OPE of the field $\mu$ as discussed in
the preceding section. The logarithmic partners of primary fields appear
precisely, if the contour of the screening charge integration gets pinched
between the two fields whose OPE is inserted.
Thus, the choice of contour together with the choice of internal
channels (due to inserted OPEs) determines which term
of the OPE is picked, either $\mu(z)\mu(0)\sim z^{1/4}(\mathbb{I}(0)+\ldots)$ 
or $\mu(z)\mu(0)\sim z^{1/4}(P(0) - 2\log(z)\mathbb{I}(0) +\ldots)$. 
So, when expanded in $\xi$, both periods,
$a(u)$ and $a_D(u)$ have asymptotics according to inserting the OPEs
$\mu(e_2)\mu(e_3)$ and $\mu(e_1)\mu(e_4)$. Keeping in mind (\ref{eq:vvev})
when inserting an OPE, we find with $e_{ij}=e_i-e_j$
\begin{eqnarray}
    a(u)&\sim&\left[e_{12}e_{13}e_{42}e_{43}\right]^{-1/4}
            \frac{e_1e_2}{e_3e_4}\left[e_{34}\Vvevss{\Phi_2(\infty)
            \mathbb{I}(e_3)P(e_4)\Phi_{-2}(0)} + \ldots\right]
            \nonumber\\[1mm]
      &\sim&\left[e_{12}e_{13}e_{42}e_{43}\right]^{-1/4}
            \frac{e_1e_2e_4}{e_3}\left[\Vvevss{\Phi_2(\infty)
            P(e_4)\Phi_{-2}(0)} + \ldots\right]
            \nonumber\\[1mm]
      &\sim&\sqrt{2u} + \ldots\,,
\end{eqnarray}
where the three-point functions evaluate trivially. Of course, the result
is the same if we had chosen the OPEs vice versa such that at the points
$e_3$ and $e_4$ we would have inserted $P(e_3)\mathbb{I}(e_4)$ instead.
In a similar fashion, we obtain
\begin{eqnarray}
    &&a_D(u)\sim\frac{1}{{\rm i}\pi}\left[e_{12}e_{13}e_{42}e_{43}\right]^{-1/4}
            \frac{e_1e_2}{e_3e_4}\left[e_{34}\Vvevss{\Phi_2(\infty)
            P(e_3)P(e_4)\Phi_{-2}(0)} + \ldots\right]
            \nonumber\\[1mm]
      && \hspace{0.5cm}\sim \frac{1}{{\rm i}\pi}
            \left[e_{12}e_{13}e_{42}e_{43}\right]^{-1/4}
            \frac{e_1e_2e_4}{e_3}\left[-2\log(e_{43})\Vvevss{\Phi_2(\infty)
            P(e_3)\Phi_{-2}(0)} + \ldots\right]
            \nonumber\\[1mm]
      &&\hspace{0.5cm}\sim\frac{{\rm i}}{\pi}\sqrt{2u}\,[\log(u) + 2\log(2) 
            + \ldots]\,.
\end{eqnarray}
Of course, other internal channels can be considered. In particular,
we may insert the OPE for $|e_1-e_3|\ll 1$ to get the behavior of the
periods for the case $u\longrightarrow\Lambda^2$. In fact, $a_D(u)$ and
$a(u)$ exchange their r\^ole since now the monopole becomes massless.
Put differently, duality in Seiberg-Witten models cooks down to
crossing symmetry in our $c=-2$ LCFT. The leading term can be read off from
$a_D(u)$ above (the OPE factors turn out to be the same up to a braiding
phase) to be proportional to $i(u-\Lambda^2)/\sqrt{2\Lambda^2}$. The relative
normalization of the logarithmic operator $\Lambda_1$ with respect to its
primary partner is fixed by considering $a_D(u)$ as
the analytic continuation of $a(u)$ via crossing symmetry yielding a 
factor of $(i\pi)^{-1}$.

There is one further BPS state which can become massless, since there is
one further zero of the discriminant
\begin{equation}
    \Delta(y^2(x))=(\det\bar{\partial}_{(j=\frac12)})^{8}=
  \left(\Vev{\prod_{i=1}^{2g+2}\Phi_{1/2}(e_i)}_{c=1}\right)^{8}
  =\prod_{j<k}(e_j-e_k)^2\,,
\end{equation}
namely $e_2\longrightarrow e_4$. This is a dyonic state with charge
$(q,g)=(-2,1)$, meaning that both, the $\alpha$ cycles as well as the
$\beta$ cycle, get pinched in this limit.
It follows that both, $a(u)$
as well as $a_D(u)$, will receive logarithmic corrections when
$u\longrightarrow -\Lambda^2$, which is well known to be the case.

Within the CFT picture, higher gauge groups as well as additional
(massive) flavors are treated in the same way. Hence, we
obtain for the $SU(2)$ case with $N_f<4$ hypermultiplets, after
simple algebra in the numerator,
\begin{eqnarray}
    \lambda_{{\rm SW}} &=& \frac{1}{2\pi{\rm i}}
    \frac{x\,{\rm d}x}{y\prod_{k=1}^{N_f}(x-m_k)}\nonumber\\
    & &{}\times\left(
    4x\prod_{k=1}^{N_f}(x-m_k) - (x-\sqrt{u})(x+\sqrt{u})\sum_{k=1}^{N_f}
    \prod_{l\neq k}(x-m_l)\right)\\
    &=& \frac{{\rm d}x}{2\pi{\rm i}}\left(( 4-N_f)\frac{x^2}{y} + N_f\frac{u}{y}
    -\sum_{k=1}^{N_f}m_k\left(\frac{x^2}{y(x-m_k)} - \frac{u}{y(x-m_k)}
    \right)\right)\,,\nonumber
\end{eqnarray}
such that we immediately can express the periods of the Seiberg-Witten form
in 4-point and 5-point functions. Using $\frac{x^2}{y(x-m_k)} =
\frac{x+m_k}{y}+\frac{m_k^2}{y(x-m_k)}$ to rewrite the last term, we
obtain
\begin{eqnarray}
    \oint\lambda_{{\rm SW}} &=& \frac{1}{2\pi{\rm i}}\left(
    (4-N_f)\Vvevss{\Phi_2(\infty)\mu(e_1)\mu(e_2)\mu(e_3)\mu(e_4)\Phi_{-2}(0)}
    \vphantom{\int}\right.\nonumber\\[1mm]
    &&+\,uN_f\Vvevss{\mu(e_1)\mu(e_2)\mu(e_3)\mu(e_4)}  
    \nonumber\\
    &&-\sum_{k=1}^{N_f}m_k\left[\vphantom{\sum}
    \Vvevss{\Phi_1(\infty)\mu(e_1)\mu(e_2)\mu(e_3)\mu(e_4)\Phi_{-1}(-m_k)}
    \right.\nonumber\\
    & &\left.\left.
       -(u-m_k^2)\Vvevss{\Phi_{-1}(\infty)\mu(e_1)\mu(e_2)\mu(e_3)\mu(e_4)
      \Phi_1(m_k)}
    \vphantom{\sum}\right]\vphantom{\int}\right)\phantom{xx}
\end{eqnarray}
as the CFT expression.
We recover hence the well know result that for all $m_k=0$ the scalar
modes have roughly the same form as in the $N_f=0$ case.
Including the charge balance at infinity, and again using $e_{ij}=
e_i-e_j$,
the above results in the following expression ($x(\cdot) = 1/M(\cdot)$
denote the inverse crossing ratios)
\begin{eqnarray}
      \oint\lambda_{{\rm SW}} &=& \left(
      \frac{(4-N_f)e_3^2}{(e_{43})^{\frac12}(e_{21})^{\frac12}}
      F_D^{(3)}({\textstyle\frac12,\frac12,-2,2,1};x(e_4),x(0),x(\infty))
      \vphantom{\sum_{k=1}^{N_f}}\right.\\
    &+&\frac{uN_f}{(e_{21})^{\frac12}(e_{43})^{\frac12}}
      \,{}_2F_1({\textstyle\frac12,\frac12;1};x(e_4))\nonumber\\
    &-& \sum_{k=1}^{N_f}\frac{m_k(e_3+m_k)}
      {(e_{21})^{\frac12}(e_{43})^{\frac12}}
      F_D^{(3)}({\textstyle\frac12,\frac12,-1,1,1};
      x(e_4),x(-m_k),x(\infty))\nonumber\\
    &+& \left.\sum_{k=1}^{N_f}\frac{m_k(u-m_k^2)}
      {(e_{21})^{\frac12}(e_{43})^{\frac12}(e_3-m_k)}
      F_D^{(3)}({\textstyle\frac12,\frac12,1,-1,1};
      x(e_4),x(m_k),x(\infty))\right)\,.\nonumber
\end{eqnarray}
Since the $F_D^{(3)}$ Lauricella functions have a negative integer
as one of the numerator parameters, they can be expanded as
polynomials in $F_1$ Appell functions, i.e.\ 5-point functions via
\begin{equation}
   F_D^{(3)}(a;b,b',b'';c;x,y,z) =
    \sum_{m=0}^{\infty}\frac{(a)_m(b')_my^m}{(1)_m(c)_m}\,
    F_1(a+m;b,b'';c+m;x,z)\,,
\end{equation}
since this expansion truncates for $b'\in\mathbb{Z}_-$.
Of course, we could have expressed this from the beginning
by only one correlation function proportional to
$F_D^{(2N_f+3)}$ of $2N_f+3$ variables, as indicated in (\ref{eq:periods}),
which is to be contrasted with the approach taken in \cite{Cappelli}.

As one further example, we consider $SU(3)$ without hypermultiplets,
where $R(Z) = \Lambda^6/(Z^3 - uZ + v)^2$ such that the resulting hyperelliptic
curve has six branch points $e_i$ and its metric $|\lambda_{{\rm SW}}|^2$
possesses three zeroes $z_j$. We get
\begin{eqnarray}
  a\lefteqn{
  \oint_{\gamma}\lambda_{{\rm SW}} = 2\Vvevs{\Phi_{2}(\infty)\mu(e_1)\ldots
     \mu(e_6)\Phi_{-1}(-\sqrt{u/3})\Phi_{-1}(0)
     \Phi_{-1}(\sqrt{u/3})}_{(\gamma)}}\nonumber\\
  &=&\prod_{i=1}^3(\partial_{e_i}M(e_i))^{\frac14}
     \prod_{i=4}^6\left(\frac{\partial_{e_i}M(e_i)}{M(e_i)^2}\right)^{\frac14}
     \prod_{j=1}^3\left(\frac{\partial_{z_j}M(z_j)}{M(z_j)^2}\right)^{-\frac12}
     \lim_{z\rightarrow\infty}\left(
     \frac{z^2\partial_{z}M(z)}{M(z)^2}\right)\nonumber\\
  &\times& F_D^{(7)}
    ({\textstyle\frac12,\frac12,\frac12,\frac12,
    -1,-1,-1,2,1};
  \nonumber\\ &&\phantom{F_D^{(7)}(}
    x(e_4),x(e_5),x(e_6),
    x(0),x({\textstyle-\sqrt{u/3}}),x({\textstyle\sqrt{u/3}}),
    x(\infty))\,,
\end{eqnarray}
with the last equality valid for $\gamma=\alpha_1\equiv C(e_2,e_3)$.
This Lauricella $D$-system for seven variables provides the complete
set of all periods. There exist more compact expressions in the
literature for this case, where the Appell function $F_4$ is involved
\cite{KlemmLerche}. However, presenting the solution in this way is more
transparent, if we view the moduli space of low-energy effective
field theory as created from string- or $M$-theory, e.g.\ as intersecting
$NS$-5 and $D$-4 branes. Then, the branch points $e_i$ and mass poles
$m_k$ are the directly given data -- they denote the endpoints of the
intersections. It remains to interpret the zeroes of the Seiberg-Witten form
within the brane picture, since they appear on equal footing with the
other singular points in our CFT approach. Moreover, this approach
suggests that BPS states from geodesic integration paths 
\cite{geodesics1,geodesics2,geodesics3} joining two
zeroes of $\lambda_{{\rm SW}}$ can be described in much the same way
as the more familiar BPS states connected to the periods. The zeroes
of $\lambda_{{\rm SW}}$ correspond to branching points in the fibration
of Calabi-Yau threefold compactifications of type II string theory, and
the corresponding BPS states are related to 2-branes ending on the
5-brane worldvolume $\mathbb{R}^4\times\Sigma$. 

Expressing the Seiberg-Witten periods in terms of
correlation functions reveals a further complication in exploring the
moduli space of low-energy effective field theories. These periods depend
only on the moduli $s_k$ and perhaps masses $m_l$. So, for the $SU(3)$
example above, the periods really depend only on two variables, $u,v$.
However, $\lambda_{\rm SW}$ in its factorized form naturally leads
to a 10-point function! The complete set of solutions of the
associated Lauricella $F_D^{(7)}$ system which covers all of $\mathbb{C}^7$ is
actually quite large, and exceeds by far the set of periods obtainable from
simple paths enclosing two of the singular points (Pochhammer paths).
As is demonstrated in \cite{Ext72}, one needs in addition at least
so-called trefoil loops which are self-intersecting contours dividing the
set of singularities into three disjunct groups.

The reason behind all this enrichment is buried in the fact that we are
dealing with a Riemann surface together with an associated metric
$\lambda_{\rm SW}$.
A detailed analysis of all these features relies on a deeper knowledge of the
analytic properties of Lauricella functions and
will be carried out in our forthcoming paper \cite{myself}.

\section{The plane versus the torus}

In this last section, we wish to make more explicit contact between the
quantities we computed for the elliptic case $g=1$, i.e.\ the torus, by
computing correlation functions on the plane with four $\mathbb{Z}_2$
branch point insertions, and well known results for the torus, parameterized
as a lattice spanned by $(1,\tau)$, the modulus of the torus. When we
considered the conformal blocks of the four-point functions
$\langle\mu(\infty)\mu(1)\mu(x)\mu(0)\rangle$, we essentially looked at the
periods of the torus. Now, in canonical normalization, we have that the
integral of the unique holomorphic form $\omega$ on the torus is given by
\begin{equation}
  \oint_\alpha\omega = 1\,,\ \ \ \ \oint_\beta\omega = \tau\,.
\end{equation}
What does this have to do with our results, which were expressed in terms
of hypergeometric functions,
\begin{equation}
  \oint_\alpha\omega = [x(1-x)]^{1/4}{}_2F_1(
  {\textstyle\frac12,\frac12,1};x)\,,\ \ \ \
  \oint_\beta\omega = [x(1-x)]^{1/4}{}_2F_1(
  {\textstyle\frac12,\frac12,1};1-x)\,?
\end{equation}
How are the torus amplitudes be related to correlation functions on the
plane?  Before we can see, how these quantities are indeed related, we first
have to briefly review the characters and torus amplitudes.

\subsection{Characters and torus amplitudes}

The theory with $c=-2$ is presumably the best understood LCFT. It is possible
to compute characters for all its irreducible representations. Besides
the vacuum representation, there are the admissible irreducible representations
given by the highest weight states created from the twist fields 
$\mu_\lambda$. Kausch \cite{Kau00} has analyzed all of these for the rational
case $n\lambda\equiv 0$ mod $1$ for an $n\in\mathbb{N}$. We will restrict
ourselves here to the case $n=2$, i.e.\ the twist fields which generate
hyperelliptic ramifications. The characters of the highest weight
representations for $h=-1/8$ and $h=3/8$ are given by
\begin{equation}
  \chi_{-1/8}(q) = \frac{\theta_{0,2}(q)}{\eta(q)}\,,\ \ \ \
  \chi_{3/8}(q) = \frac{\theta_{2,2}(q)}{\eta(q)}\,,
\end{equation}
where the Jacobi-Riemann theta-functions and the Dedekind eta-function are
defines as
\begin{equation}\label{eq:mytheta}
  \theta_{\lambda,k}(q) = \sum_{n\in\mathbb{Z}}q^{(2kn+\lambda)^2/4k}\,,\ \ \ \
  \eta(q) = q^{1/24}\prod_{n=1}^\infty(1-q^n)\,.
\end{equation}
There are two more irreducible representations, namely for $h=0$ and $h=1$,
with characters
\begin{equation}
  \chi_0(q) = \frac{1}{2\eta(q)}(\theta_{1,2}(q) - \eta(q)^3)\,,\ \ \ \
  \chi_1(q) = \frac{1}{2\eta(q)}(\theta_{1,2}(q) + \eta(q)^3)\,.
\end{equation}
The problem is that these two characters do not have a good homogeneous
transformation behavior under the modular group. More specifically,
the transformation $S:\tau\rightarrow-1/\tau$ will map these characters
into functions, where $\tau$ appears directly as a prefactor, and not only
via $q=\exp(2\pi{\rm i}\tau)$. In other words, the $S$-matrix has entries
which are not all constant with respect to $\tau$. We already know, however,
that the vacuum representation is an irreducible sub-representation of a 
larger, indecomposable, representation ${\cal R}$. The point is that
the character of the full indecomposable representation is well behaved,
\begin{equation}
  \chi_{{\cal R}}(q) = \frac{\theta_{1,2}(q)}{\eta(q)}\,.
\end{equation}
The modular transforms of the vacuum-character, however, can be given in terms
of an $S$-matrix with constant coefficients, if we enlarge the set of
characters by functions of the form
\begin{equation}
  \tilde\chi_0(q) = \frac{\log(q)}{2\pi{\rm i}}\eta(q)^2\,,
\end{equation}
where we choose form $\log(q)$ the branch that coincides with $2\pi{\rm i}\tau$.
This function cannot be interpreted as character, but it still is a valid
torus amplitude. This is a manifestation of a more general fact, namely that
the space of torus amplitudes and the space of characters are, as vector
spaces, not any longer isomorphic, if indecomposable representations have
to be taken into account, as is the case in LCFT.

\subsection{Periods and Torus Amplitudes}

We now come to the point where we can compare our results for the periods
on the torus, obtained by a computation on a ramified covering of the sphere,
with the well-known results in terms of elliptic functions, depending on the
modulus $\tau$ of the torus. We thus will make some use of the theory
of elliptic functions in order to compute torus amplitudes directly
on the complex plane. Actually, we will do the computation on the two-sheeted
covering of $\mathbb{CP}^1$ with four branch points $e_1,\ldots,e_4$.
We will then translate the result to expressions in the modular parameter
$\tau$ with the help of some identities for elliptic functions.

We recall that a torus is an elliptic curve defined by an equation
\begin{equation}
  y^2 = (z-e_1)(z-e_2)(z-e_3)(z-e_4)\,,
\end{equation}
where infinity is not taken as a branch point. We note that the branch points
are related to the modular parameter $\tau$ in a non-trivial manner. 
Due to conformal invariance, we can fix three of the four branch points
to arbitrarily chosen coordinates. However, we will not directly make use
of this yet, but instead assume only that $e_4=\infty$ and $e_2+e_3+e_1=0$. 
It is easy to see that this is possible without loss of generality.
Then, elementary
symmetric polynomials in the $e_i$ can be expressed in terms of modular
functions $g_2(\tau)$ and $g_3(\tau)$, which essentially are the Eisenstein 
series $E_4(\tau)$ and $E_6(\tau)$, respectively. This holds more generally,
but is particularly easy in this setting, where we then have $y^2=4z^3 - g_2z
-g_3$. Moreover, the so-called discriminant of the elliptic curve, which 
reads
\begin{equation}
  \Delta = \prod_{i<j}(e_i-e_j)^2
\end{equation}
up to an irrelevant numerical constant, 
is nothing else than the modular invariant $\Delta(\tau)=\eta(\tau)^{24}$.
Here, $\eta(\tau)$ is the Dedekind $\eta$-function defined above.

The important point is that there is no simple relation between the
branch points and the modular parameter $\tau$. However, one can make
use of a so-called uniformizing variable $u$ which allows to define 
the elliptic curve in terms of an elliptic function, namely the
Weierstrass' function $\wp$. One has
\begin{equation}
  \wp'(u) = 4\wp^3(u) - g_2\wp(u) - g_3\,.
\end{equation}
This relation as well as some relations involving standard 
$\vartheta$-functions will be helpful later on.

We first will do an indirect computation: On the branched covering of 
$\mathbb{CP}^1$, the periods of the
corresponding torus can be computed by contour integrations in the following
way: Firstly, we choose a basis of homology cycles, where $\alpha$ encircles
$e_2$ and $e_3$, and $\beta$ encircles $e_1$ and $e_3$. (This choice is
conventional, and assumes that the branch cuts run between $e_2$ and $e_3$,
and between $e_1$ and $e_4$, respectively.) Therefore,
\begin{eqnarray}
  \pi_{\alpha} &=& \oint_{\alpha}\frac{{\rm d}z}{y}\ =\ 
                   \frac{\sqrt{2}}{2\pi}\int_{e_2}^{e_3}\frac{{\rm d}z}{y}\ =\
                   \frac{\sqrt{2}}{2\pi}\int_{e_2}^{e_3}{\rm d}z
                   \prod_{i=1}^4(e_i-z)^{-1/2}\,,\\[1mm]
  \pi_{\beta}  &=& \oint_{\beta}\frac{{\rm d}z}{y}\ =\
                   \frac{\sqrt{2}}{2\pi}\int_{e_1}^{e_3}\frac{{\rm d}z}{y}\ =\
                   \frac{\sqrt{2}}{2\pi}\int_{e_1}^{e_3}{\rm d}z
                   \prod_{i=1}^4(e_i-z)^{-1/2}\,.
\end{eqnarray}
The correct proportionality factors connecting the contour integrations to
line integrations follow from the theory of hypergeometric integrals.
The above integrals look very much like Feigin-Fuks screening integrals
in a free field representation of a CFT. Indeed, they are just that for
the CFT of 1-differentials, the ghost systems with $c=-2$ which we use
throughout this paper. More precisely,
we have, with $\mu$ the $\mathbb{Z}_2$ twist field of conformal weight 
$h=-1/8$ simulating a branch point, that
\begin{equation}
  \frac{\sqrt{2}}{2\pi}\int_{e_k}^{e_l}{\rm d}z\prod_{i=1}^4(e_i-z)^{-1/2} =
  \prod_{i<j}(e_i-e_j)^{-1/4}
  \langle\mu(e_1)\mu(e_2)\mu(e_3)\mu(e_4)\rangle_{C_{(e_k,e_l)}}\,,
\end{equation}
where we have to divide the correlator by its free part such that only the
screening integration remains. It is clear that different line integrations
lead to different linear combinations of conformal blocks, indicated here
by the notation $C_{(e_k,e_l)}$ for a contour encircling only the branch
points $e_k$ and $e_l$. Thus, we can write
\begin{eqnarray}
  \pi_{\alpha} &=& \Delta^{-1/8}
                   \langle\mu(e_1)\mu(e_2)\mu(e_3)\mu(e_4)
                   \rangle_{(\alpha)}\,,\\[1mm]
  \pi_{\beta}  &=& \Delta^{-1/8}
                   \langle\mu(e_1)\mu(e_2)\mu(e_3)\mu(e_4)
                   \rangle_{(\beta)}\,.
\end{eqnarray}

Now, with our choice of the homology basis, the periods of the torus
should simply be $\pi_{\alpha}\sim 1$ and $\pi_{\beta}\sim \tau$. There
is one subtlety to be taken care of, namely that we have computed the periods
not in a flat metric, but in a singular metric of the branched covering of 
the complex plane. The periods depend on the metric due to the conformal 
anomaly. The difference between these two metrics is well known and
amounts to $\prod_{i<j}(e_i-e_j)^{-1/12}=\Delta^{-1/24}=\eta^{-1}$. In the
literature, this is often called the Liouville factor. Putting all together,
we find the result
\begin{eqnarray}
  \Delta^{-1/8}\Delta^{1/24}\pi_{\alpha} &=& \eta^2\cdot 1\ =\
               \langle\mu(e_1)\mu(e_2)\mu(e_3)\mu(e_4)
               \rangle_{(\alpha)}\,,\\
  \Delta^{-1/8}\Delta^{1/24}\pi_{\beta} &=& \eta^2\cdot \tau\ =\
               \langle\mu(e_1)\mu(e_2)\mu(e_3)\mu(e_4)
               \rangle_{(\beta)}\,.
\end{eqnarray}
Therefore, we claim that the conformal blocks $\langle 1\rangle_{\gamma}$
on the torus are given by $\eta^2(\tau)$ for $\gamma=\alpha$, and by
$\eta^2(\tau)\tau$ for $\gamma=\beta$.

Now we rederive this result by a more direct computation: We could continue and 
evaluate the correlators. It is well known that
these turn out to be proportional to the hypergeometric system
${}_2F_1(\frac12,\frac12;1;x)$ with $x$ the anharmonic ratio of the four
branch points. In principle, it is possible to express the resulting
function in $x$ in terms of $\tau$ using the relation
$x=\kappa^2(\tau)=(\vartheta_2(0|\tau)/\vartheta_3(0|\tau))^4$, 
where $\vartheta_i(v|\tau)$, $i=1\ldots 4$,
denote the standard Jacobi $\vartheta$-functions. These are given as
special cases of the Hermite $\Theta$-function
\begin{equation}
  \Theta_{\mu,\nu}(v|\tau) = \sum_{n\in\mathbb{Z}}\exp\left[
  2\pi{\rm i}\tau{\textstyle\frac12}(n+{\textstyle\frac12}\mu)^2 
  + 2\pi{\rm i}v(n+{\textstyle\frac12}\mu) + \pi{\rm i}n\nu\right]\,,
\end{equation}
namely $\vartheta_1(v|\tau)={\rm i}\Theta_{-1,1}(v|\tau)$, 
$\vartheta_2(v|\tau)=\Theta_{-1,0}(v|\tau)$,
$\vartheta_3(v|\tau)=\Theta_{0,0}(v|\tau)$, and 
$\vartheta_4(v|\tau)=\Theta_{0,1}(v|\tau)$.
The inverse relation is
\begin{equation}\label{eq:tau}
  \tau = {\rm i}\frac{{}_2F_1(\frac12,\frac12;1;1-x)}{
                      {}_2F_1(\frac12,\frac12;1;x)}\,.
\end{equation}
Relations between 
expressions in the modular parameter and expressions in terms of branch 
points are often called (generalized) Thomae's formul\ae. We can make this
more manifest by remembering that the complete elliptic integral of the
first kind can be expressed in terms of a hypergeometric system, namely
\begin{equation}
  K(x) = {}_2F_1({\textstyle \frac12,\frac12;1;x})\,,\ \ \ \ 
  K'(x)={}_2F_1({\textstyle \frac12,\frac12;1;1-x})\,.
\end{equation}
On the other hand, one knows from the theory of elliptic curves that
\begin{equation}\label{eq:e1e3}
  \begin{array}{rclcr}
  K(x) &=& \phantom{-{\rm i}}(e_1-e_3)^{\frac12}\omega          &=& 
           \frac{\pi}{2}\vartheta_3^2(0|\tau)\,,\\[2mm]
  K'(x)&=& -{\rm i}(e_1-e_3)^{\frac12}\omega' &=& 
           -{\rm i}\frac{\pi}{2}\frac{\omega'}{\omega}\vartheta_3^2(0|\tau)\,, 
  \end{array}
\end{equation}
where $\omega$ and $\omega'$ are the two periods of the elliptic curve.
Now, the four-point function can be explicitly evaluated up to an overall
constant as
\begin{equation}
  \left\langle\mu(\infty)\mu(e_1)\mu(e_2)\mu(e_3)\right\rangle =
  -\pi\left(\frac{(e_1-e_2)(e_2-e_3)}{(e_1-e_3)}\right)^{\frac14}F(x)\,,
\end{equation}
where $F(x)$ is a linear combination in $K(x)$ and $K'(x)$. Luckily,
differences of branch points can be expressed in terms of $\vartheta$-functions
as well. In addition to (\ref{eq:e1e3}), we need
\begin{equation}
  (e_2-e_3)^{\frac12} = \frac{\pi}{2\omega}\vartheta_2^2(0|\tau)\,,\ \ \ \
  (e_1-e_2)^{\frac12} = \frac{\pi}{2\omega}\vartheta_4^2(0|\tau)\,.
\end{equation}
Thus, the prefactor results in
\begin{equation}
  \left(\frac{(e_1-e_2)(e_2-e_3)}{(e_1-e_3)}\right)^{\frac14} =
  \sqrt{\frac{\pi}{2\omega}}\frac{\vartheta_2\vartheta_4}{\vartheta_3} =
  \sqrt{\frac{\pi}{2\omega}}\frac{\vartheta_2\vartheta_3\vartheta_4}{
     \vartheta_3^2}\,,
\end{equation}
where we abbreviate $\vartheta_\alpha=\vartheta_\alpha(0|\tau)$.
Making use of the relation
\begin{equation}
  \Delta^{\frac14} = \frac{\pi^3}{4\omega^3}\left(\vartheta_2\vartheta_3
  \vartheta_4\right)^2\,,
\end{equation}
and plugging in the above formula for $K(x)$ and $K'(x)$, we arrive at
the statement
\begin{equation}
  \left\langle\mu(\infty)\mu(e_1)\mu(e_2)\mu(e_3)\right\rangle =
    -\pi\frac{\sqrt{2}}{\pi}\eta^3\left\{\begin{array}{r}
      \omega\phantom{'}\,,\\
      ({\rm -i})\omega'\,.
    \end{array}
  \right.
\end{equation}
This is, up to the above mentioned Liouville factor 
$\prod_{i<j}(e_i-e_j)^{c/12}$ and an overall scale 
equivalent to our first computation.
As a further example we give without computation the result
\begin{equation}
  \left\langle\mu(\infty)\mu(e_1)\mu(e_2)\mu(e_3)V_{1}(z)\right\rangle =
  -\frac{1}{\eta^6}\sqrt{\frac{\pi^3}{\omega^3}}\frac{\vartheta_1^4(v|\tau)}{
  \vartheta_1(2v|\tau)}\,,
\end{equation}
where we have not yet divided out the Liouville factor, and where 
$v$ is related to $z$ via $z=\wp(v)$. The field $V_1(z)$ denotes here
a vertex operator of charge $q=1$, such that its conformal weight is
$h(q)=\frac12(q^2+q)=1$. Since the total charge balance is already satisfied,
no screening charge integrations are necessary.

\subsection{Plane correlators and characters}

Finally, we will make some more direct connections of all these quantities
and with the character functions for the $c=-2$ rational LCFT. The key to this
is the relation (\ref{eq:tau}) with its inverse, the elliptic modular function
$x(\tau)=\kappa^2(\tau)$. If one wishes, $\kappa^2(\tau)$ can be expressed in
terms of the Dedekind eta-function as
\begin{equation}
  \kappa^2(\tau) = 16\frac{\eta(2\tau)^{16}\eta(\tau/2)^8}{\eta(\tau)^{24}}\,.
\end{equation}
Obviously, the relation (\ref{eq:tau}) is equivalent to the
quotient of our two conformal blocks of four twist fields,
\begin{eqnarray}
  \tau &=& \frac{\omega}{\omega'} = 
  {\rm i}\frac{\bra\mu(\infty)\mu(1)\mu(x)\mu(0)\ket_\beta}
  {\bra\mu(\infty)\mu(1)\mu(x)\mu(0)\ket_\alpha}\\[1mm]
  &=& {\rm i}\left(\log(x)\! -\! 4\log(2)\! -\! \frac12\, x\! 
-\! \frac{13}{64}\,x^2
     \! -\!\frac{23}{192}\,x^3\! -\! \frac{2701}{32768}\,x^4\! -\! \frac{5067}{81920}\,x^5 -
      \ldots\right)\,.\nonumber
\end{eqnarray}
In fact, one finds the remarkable results that not only the quotient of the two
conformal blocks can be expressed in modular quantities, but also the
individual hypergeometric functions and the two conformal blocks themselves.
First of all, one finds that
\begin{equation}
  {}_2F_1({\textstyle\frac{1}{2},\frac{1}{2},1};\kappa^2) = (\vartheta_3)^2\,,
  \ \ \ \
  {}_2F_1({\textstyle\frac{1}{2},\frac{1}{2},1};1-\kappa^2) = 
  -{\rm i}\tau(\vartheta_3)^2\,,
\end{equation}
as functions of $\tau$. The prefactor of the two conformal blocks, however,
does not have a simple expression in terms of $\vartheta_\alpha$ functions.
It is now useful, to switch to the characters of the $c=-2$ theory. One has
to note one subtlety here. Elliptic functions as the ones introduced above,
are typically defined in the half-period $\tau$ with Fourier-expansions
around $\tau=+{\rm i}\infty$ in the variable $\tilde q=\exp(\pi{\rm i}\tau)
=q^{1/2}$, although this variable is denoted often with $q$ in the literature,
causing considerable confusion. To start with, we have explicitly
\begin{eqnarray}
  \vartheta_3 &=& 
    1 + 2\sum_{n=1}^\infty \tilde q^{n^2}= \theta_{0,2}+\theta_{2,2}\,,\\
  \vartheta_2 &=& 2\sum_{n=0}^\infty\tilde q^{(n+1/2)^2} = \theta_{1,2}
\end{eqnarray}
as functions of $\tau$. It is worth noting that the characters of the
irreducible representations for the twist fields appear only in their sum.
The difference of these characters is also a meaningful quantity, namely
\begin{equation}
  \theta_{0,2}-\theta_{2,2} = \frac{\eta^2(\tau/2)}{\eta(\tau)} = 
  \vartheta_4\ \ \ \
  \textrm{or}\ \ \ \ \chi_{-1/8}-\chi_{3/8}=\left(\frac{\eta(\tau/2)}
  {\eta(\tau)}\right)^2\,.
\end{equation}
With this in mind, we find the relations
\begin{eqnarray}
  {}_2F_1({\textstyle\frac12,\frac12,1;}\kappa^2) &=&
  (\theta_{0,2}+\theta_{2,2})^2 = \left(\eta\,(\chi_{-1/8}+\chi_{3/8})\right)^2
  \,,\\[1mm]
  \left(\kappa^2(1-\kappa^2)\right)^{1/4} &=& 2\,\frac{\theta_{1,2}(\theta_{0,2}
  -\theta_{2,2})}{(\theta_{0,2}+\theta_{2,2})^2} = 2\chi_{{\cal R}}\,
  \frac{\chi_{-1/8}-\chi_{3/8}}{(\chi_{-1/8}+\chi_{3/8})^2}\,,
\end{eqnarray}
The latter is the prefactor of the conformal blocks. But it is also a 
correlation function in its own right, namely the function
\begin{eqnarray}
  & & \langle\sigma(\infty)\mu(1)\mu(\kappa^2)\mu(0)\rangle\ =\ 
  \left[\kappa^2(1-\kappa^2)\right]^{1/4}\\
  & &\phantom{xxx}\ =\ 
  2\tilde q^{1/4}\left(1 - 6\tilde q + 21\tilde q^2 - 62\tilde q^3 
      - 162\tilde q^4 -384\tilde q^5 + 855\tilde q^6 + \ldots\right)
  \,,\nonumber
\end{eqnarray}
where $\sigma(z)=\Phi_{-1/2}(z)$. Note that the charge balance is
automatically fulfilled for this correlator, so no screening current integration
is necessary. Collecting everything
leads to the surprising compact result
\begin{eqnarray}
  \bra\mu\mu\mu\mu\ket_\alpha &=& 2\chi_{{\cal R}}(\tau)\eta^2(\tau/2)\,,\\[1mm]
  \bra\mu\mu\mu\mu\ket_\beta &=& 2\chi_{{\cal R}}(\tau)\tilde\chi_0(\tau/2)\,.
\end{eqnarray}
Up to a factor involving the Dedekind eta-function taken at the half-period,
the conformal blocks turns out to be directly related to the character of
the $h=0$ sector of the theory! We might have expected something like this,
since the four-point function on the sphere should correspond to a zero-point
function on the torus. Note that $\eta^2(\tau/2)=\chi_0(\tau/2)-
\chi_1(\tau/2)$. We pay a small price for using the picture of a ramified
covering of the sphere. The metric or Liouville factor actually enters
with the half-period instead of the full period reflecting the fact that
there is an additional $\mathbb{Z}_2$ symmetry. The realization of a torus
via a ramified covering introduces a double-valuedness which is not present,
when the torus is realized as $\mathbb{C}/\Lambda$ with $\Lambda$ a lattice
spanned by $\mathbb{Z}\omega\oplus\mathbb{Z}\omega'$. The period or modulus
$\tau$ can be
expressed solely in terms of the characters of the representations, which
are part of the indecomposable $h=0$ representation ${\cal R}$, namely
\begin{equation}
  \tau = \frac{1}{{\rm i}\pi}\frac{\tilde\chi_0}{\chi_{{\cal R}}} = 
  \frac{1}{{\rm i}\pi}\frac{\tilde\chi_0}{\chi_0-\chi_1}\,.
\end{equation}

We complete this brief discussion by presenting the other possible
four-point correlators involving twist fields. Since we need at least on
$\mu$ field, as the excited twist $\sigma$ is degenerate of level four,
the remaining two cases are, up to permutations,
\begin{eqnarray}
  \bra\sigma(\infty)\sigma(1)\mu(x)\mu(0)\ket &=& x[x(1-x)]^{-3/4}
  {}_2F_1({\textstyle-\frac12,\frac12,1};x)\,,\\[1mm]
  \bra\sigma(\infty)\sigma(1)\mu(x)\sigma(0)\ket &=& [x(1-x)]^{-1/4}\nonumber\\
  &=& \frac12\frac{(\chi_{-1/8}+\chi_{3/8})^2}{\chi_{{\cal R}}(
      \chi_{-1/8}-\chi_{3/8})}\,,
\end{eqnarray}
where $\frac{\pi}{2}{}_2F_1(-\frac12,\frac12,1;\kappa^2)=E(\kappa^2)$ is
the complete elliptic integral of the second kind.
All these functions are related to the vacuum character. They differ by
a metric factor which can be expressed in terms of the Dedekind eta-function.
If we define the symbol $(n)=\eta(n\tau/2)=\eta(q^{n/2})$, then we can write all
the different twist field four-point functions as
\begin{eqnarray}
  \bra\mu\mu\mu\mu\ket &=& \chi_{{\cal R}}\,(1)^2\,,\\[1mm]
  \bra\sigma\mu\mu\mu\ket &=& \chi_{{\cal R}}\,\frac{(1)^6(4)^4}{(2)^{10}}\,,\\[1mm]
  \bra\sigma\sigma\mu\mu\ket &=& \chi_{{\cal R}}\,\frac{(2)^4E_2(q^2)}{
    (1)^6}\,,\\[1mm]
  \bra\sigma\sigma\sigma\mu\ket &=& \chi_{{\cal R}}\,\frac{(2)^{11}}{(1)^4(4)^7}
  \,,
\end{eqnarray}
where we only give the conformal blocks for the $\alpha$ homology cycle and
where we have skipped some irrelevant numerical factors. The third case,
involving two excited twists, leads to the complete elliptic integral of the
second kind. Expressing it in terms of the Jacobi theta-functions leads to
the given result, where $E_2$ is the second Eisenstein series,
$E_2(q) = 1 - 24\sum_{k\in\mathbb{N}}\sigma_1(k)q^{2k}$ with
$\sigma_1(k)$ the sum of the divisors of $k$.

It is possible to generalize such connections between four-point
amplitudes on the sphere and torus vacuum amplitudes. The tool for this
is uniformization. In case that one field in the four-point functions is
degenerate of level two, we know that the four-point amplitude has to
satisfy a second-order differential equation. Thus, the four-point
amplitude will be related to a hypergeometric functions ${}_2F_1(a,b,c;x)$.
More precisely, if $\Phi_{h_0}$ is degenerate of level two, we can bring
any four-point function involving this field into the form
\begin{eqnarray}
  \bra\Phi_{h_3}(\infty)\Phi_{h_2}(1)\Phi_{h_1}(0)\Phi_{h_0}(z)\ket
  &=& z^{p+\mu_{01}}(1-z)^{q+\mu_{20}}F(z)\,,\\[1mm]
  \mu_{ij} &=& (h_0+h_1+h_2+h_3)/3-h_i-h_j\,,\nonumber\\[1mm]
  p &=& {\textstyle\frac16 - \frac23h_0-\mu_{01}-\frac16\sqrt{r_1}}
        \,,\nonumber\\[1mm]
  q &=& {\textstyle \frac16 - \frac23h_0-\mu_{01}-\frac16\sqrt{r_2}}
        \,,\nonumber\\[1mm]
  r_i &=& {\textstyle 1 - 8h_0 + 16h_0^2 + 48h_ih_0 + 24h_i}\,.\nonumber
\end{eqnarray}
The remaining function $F(z)$ then is a solution of the hypergeometric 
system ${}_2F_1(a,b;c;z)$ given by
\begin{eqnarray}
  0 &=& \left(z(1-z)\partial_z^2 + [c - (a+b+1)z]\partial_z - 
ab\right)F(z)\,,\\[1mm]
  a &=& {\textstyle\frac12-\frac16\sqrt{r_1}-\frac16\sqrt{r_2}
        -\frac16\sqrt{r_3}}\,,\nonumber\\[1mm]
  b &=& {\textstyle\frac12-\frac16\sqrt{r_1}-\frac16\sqrt{r_2}
        +\frac16\sqrt{r_3}}\,,\nonumber\\[1mm]
  c &=& {\textstyle 1 - \frac13\sqrt{r_1}}\,.\nonumber
\end{eqnarray}
The general solution is then a linear combination of the two
linearly independent solutions or conformal blocks ${}_2F_1(a,b;c;z)$ and
$z^{1-c}{}_2F_1(a-c+1,b-c+1;2-c;z)$. Which linear combination one has to take
is determined by the requirement that the full four-point function 
involving holomorphic and anti-holomorphic dependencies must be
single-valued to represent a physical observable quantity.

Uniformization of this case can now be done with the help of a beautiful
formula by Wirtinger, namely
\begin{equation}
  {\textstyle\frac12}\frac{\Gamma(2b)\Gamma(c-b)}{\pi^{2b}\Gamma(c)}
  {}_2F_1(a,b,c;\kappa^2(\tau))= \left[\vartheta_3(0|\tau)\right]^{4b}
  \int_0^{1/2}\Phi(u,\tau){\rm d}u\,,
\end{equation}
where the line integral is valid for $\RE c > \RE b > 0$, defining a
one-valued function regular in the half-plane $\IM \tau>0$. In all other
cases, the line integral must be replaced by a contour integration. The
function $\Phi(u,\tau)$ is entirely defined in terms of the Jacobi 
theta-functions,
\begin{equation}
  \Phi(u,\tau) = 
  \left[\frac{\vartheta_1(u|\tau)}{\vartheta'_1(0|\tau)}\right]^{2b-1}
  \left[\frac{\vartheta_2(u|\tau)}{\vartheta_2(0|\tau)}\right]^{2(c-b)-1}
  \left[\frac{\vartheta_3(u|\tau)}{\vartheta_3(0|\tau)}\right]^{1-2a}
  \left[\frac{\vartheta_4(u|\tau)}{\vartheta_4(0|\tau)}\right]^{1-2(c-a)}\!\!,
\end{equation}
where $\vartheta_1'(v|\tau)=\partial_v\vartheta_1(v|\tau)$.

As a small example, we look at the Ising model with $c=c_{4,3}=1/2$. The field
of conformal weight $h=h_{2,1}=1/2$ is degenerate of weight two. Its
four-point function is a rational function, namely
\begin{equation}
  \bra\Psi_{2,1}(\infty)\Psi_{2,1}(1)\Psi_{2,1}(x)\Psi_{2,1}(0)\ket\! =\!
  [x(1\!-\!x)]^{-1}{}_2F_1({\textstyle -2,-\frac13,-\frac23};x) =
  \frac{1\!-\!x\!+\!x^2}{x^2(1\!-\!x)^2}\,.
\end{equation}
Expanding this for $x=\kappa^2(\tau)$ results in the series
\begin{eqnarray} & &
  \bra\Psi_{2,1}(\infty)\Psi_{2,1}(1)\Psi_{2,1}(\kappa^2)\Psi_{2,1}(0)\ket\ =
  \\[1mm]
  & & \phantom{xxxx}
  \frac{1}{16\tilde q}\left(1 + 8\tilde q + 276\tilde q^2 + 2048\tilde q^3
  + 11202\tilde q^4 + 49152\tilde q^5 + 184024\tilde q^6 + \ldots\right)
  \nonumber
\end{eqnarray}
in the variable $\tilde q=q^{1/2}$. This is a known series, namely the
McKay-Thompson series of class 4A for the Monster, which is the character
of the extremal vertex operator algebra of rank 12. We find it most
astonishing that this function appears within the Ising model!

The Rising model admits three irreducible
representations, with conformal weight $h_{1,1}=0$, $h_{2,1}=1/2$ and
$h_{1,2}=1/16$, respectively. The characters are all given in terms of the
modular functions
\begin{eqnarray}
  K_{p,p',r,s} &=& \frac{1}{\eta(q)}
  \sum_{n\in\mathbb{Z}}q^{(2pp'n+pr-p's)^2/4pp'} = 
  \frac{\theta_{pr-p's,pp'}(q)}{\eta(q)}\,,\\
  \chi_{p,p',r,s} &=& K_{p,p',r,s}-K_{p,p',r,-s} =
  \frac{\theta_{pr-p's,pp'}(q)-\theta_{pr+p's}(q)}{\eta(q)}\,,
\end{eqnarray}
with the functions defined in (\ref{eq:mytheta}). These expressions yield
the characters for all representations from the conformal grid $h_{r,s}$
of all minimal models $c_{p,p'}$. Expressing these in
$\tilde q$, one finds the remarkable identity
\begin{equation}
  \bra\Psi_{2,1}(\infty)\Psi_{2,1}(1)\Psi_{2,1}(\kappa^2)\Psi_{2,1}(0)\ket =
  (\chi_{4,3,1,1}+\chi_{4,3,2,1})^{24} - 1
\end{equation}
up to an irrelevant numerical factor. But the most remarkable property of
this four-point function is that it is closely related to the modular invariant
$J(\tau)$, namely
\begin{equation}
  J(\tau)=\frac{4}{27}\frac{(1-\kappa^2+\kappa^4)^3}{\kappa^4(1-\kappa^2)^2}
  = \frac{4}{27}(1-\kappa^2+\kappa^4)^2
  \bra\Psi_{2,1}(\infty)\Psi_{2,1}(1)\Psi_{2,1}(\kappa^2)\Psi_{2,1}(0)\ket\,.
\end{equation}

Of course, we do not get simple relations between the four-point functions
of a given CFT and its torus vacuum-amplitudes, as we found in the case of
the special theory with $c=-2$. The reason is that the branched covering of
the Riemann sphere, generated by the fields in an arbitrary four-point
functions, is not necessarily a torus. In fact, it almost never is. The reader
should keep in mind that even in the case that (at least) one of the fields
is degenerate of level two such that one screening integration has to be
performed, does not imply that we have the geometry of a torus. On the
contrary, except for $c=-2$, where the contours can indeed be chosen as
homology cycles, we are almost always forced to use Pochhammer double loops
to get a contour which closes. However, what the uniformization tells us is
that we can view the geometries generated by the four fields in the 
correlator as a torus-like branched covering of some non-trivial Riemann
surface. Let us make this a bit more precise by looking at another
four-point function in the Ising model, namely at the correlator of the
field $\Phi_{1,2}$ with $h=h_{1,2}=1/16$ which is also degenerate of level
two. We find
\begin{equation}
  \langle\Phi_{1,2}(\infty)\Phi_{1,2}(1)\Phi_{1,2}(x)\Phi_{1,2}(0)\rangle
  = \left[x(1-x)\right]^{-1/8}{}_2F_1({\textstyle-\frac14,\frac14,\frac12};x)
  \,.
\end{equation}
Making use of one of the elementary relations for hypergeometric functions,
this can be expressed as
\begin{equation}
  \langle\Phi_{1,2}(\infty)\Phi_{1,2}(1)\Phi_{1,2}(\kappa^2)
  \Phi_{1,2}(0)\rangle =
  \frac{1}{\sqrt{2}}
  \left[\kappa^2(1-\kappa^2)\right]^{-1/8}
  \sqrt{1 + \sqrt{1-\kappa^2}}\,.
\end{equation}
Comparing this with the expansions in the elliptic modulus $\kappa^2$ for
the correlation functions of the $c=-2$ theory, we see that the spin
four-point correlation function of the Ising model goes with $(\kappa^2)^{1/8}$,
while all twist field correlators in the $c=-2$ model go with
$(\kappa^2)^{1/4}$. So, there is an {\slshape additional\/} multi-valuedness,
an additional branching.

\section{Conclusion}

We have collected here various issues concerning logarithmic conformal
field theory. We concentrated on the best known example of a LCFT, namely
the rational theory with $c=c_{2,1}=-2$. The reason for this is that we
were mainly interested in the geometrical meaning of the logarithmic
operators. This particular conformal field theory is different to other
conformal field theories such as the minimal models, in the sense that its
fields have an immediate geometrical meaning, such as branch points or
poles, inserted in the complex plane. Thus, field insertion very directly
change the geometry from the complex plane to a ramified covering of it.
This works much more naturally than in other CFTs, where the ramified 
coverings are defined only locally and screening contours do not close, when
one attempts to choose simple loops encircling two field insertions. Instead,
in the generic CFT case, one has to take Pochhammer double loops.

In the LCFT setting discussed here, screening contours can be chosen as
ordinary homology cycles. But this implies that logarithmic operators must
appear. Considering branch points as certain vertex operators, and then
viewing these vertex operators as dynamical objects we can move around,
leads to a possible degenerate case. When we let two branch points run
into each other, a homology cycle might get pinched between them. If the
two branch points have monodromies, which cancel each other, the pinched
cycle leads to a defect, since it cannot be closed on the same sheet. 
Deforming the contour leaves one with a small line integral of a simple
pole, giving rise to the logarithmic divergences one observes in LCFT.

After having interpreted the origin of logarithms in a geometrical way, we 
consider two settings where moving around of branch points plays an important
role. Firstly, we look at the Seiberg-Witten solutions of supersymmetric
low energy effective field theories. It is a well known fact that the
periods of a natural meromorphic one-form, defined on a hyperelliptic
surface, either vanish or exhibit a logarithmic behavior, when certain
homology cycles shrink to zero size. This behavior is, of course and naturally,
precisely recovered, if one computes the periods in terms of LCFT
correlators of the $c=-2$ system of analytic one-differentials.

The second interesting topic we wished to raise was a direct comparison
between quantities one can compute directly on the torus, and plane
correlations functions with four $\mathbb{Z}_2$ twist fields inserted to
simulate the torus. We have tried to make some of these connections very
explicit. However, we believe that this view point is rather uncommon,
and so, not much is yet known about such relations. In particular, almost
nothing can be said for similar considerations on other CFTs. 
However, we are convinced that such connections between ramified coverings of
the plane and non-trivial Riemann surfaces, in particular the torus, must
generally exist. We leave the exploration of this issue to future work.

Last but not least, we hope that this small contribution, attempting to 
discuss various topics in physics and to make use of mathematical structures 
from many fields, combining some things in slightly uncommon ways,
is a bit in the spirit of Ian Kogan, circumnavigating
a very small part of physics.

\section*{Acknowledgments} 
Throughout the years, I have enjoyed many fruitful discussions with
and comments from Nikolas Akerblom, Ralph Blumenhagen, Andreas Bredthauer,
John Cardy, Holger Eberle, Matthias Gaberdiel,
Amihay Hanani, Andreas Honecker, Marco Krohn, Neil Lambert, Alex Nichols,
Werner Nahm, Klaus Osterloh, Hubert Saleur, Julia Voelskow, Gerard Watts, 
Peter West, John Wheater, and, of course, Ian Kogan. 

\appendix
\renewcommand{\theequation}{A.\arabic{equation}}
\setcounter{equation}{0}
\section{Structure Constants}

structure constants of the $c_{p,1}$ LCFTs. These are needed, if
correlation functions are decomposed into 2- 3- and 4-point functions.
The case important for this paper is $p=2$.
One starts from an ansatz for the OPE,
\begin{equation}
  \Phi_q(z)\Phi_{q'}(w) = \sum_{q''}C_{q,q'}^{q''}\,
    (z-w)^{h(q'')-h(q)-h(q')}\,\Phi_{q''}(w) + \ldots
\end{equation}
where we omitted logarithmic contributions and descendants and where
$q,q',q''$ all are admissible charges according to (\ref{eq:admissible}). The 
local primary fields decompose into chiral vertex operators with
coefficients determined by the braid matrices of the BRST-invariant vertex 
operators. Depending on whether the local primary fields are chiral or
not, we put $\chi=1,0$ respectively and have
\begin{eqnarray}
    {\displaystyle\left({\cal D}_{(n,n'),(m,m')}^{(l,l')}\right)^{1+\chi}} &=&
    {\displaystyle \frac{N_{(l,l')(l,l')}^{(1,1)}}{N_{(n,n')(n,n')}^{(1,1)}
    N_{(m,m')(m,m')}^{(1,1)}}
    \left(\Delta_{n,m}^{l}(x)\Delta_{n',m'}^{l'}(x')\right)^{1+\chi}}
    \,,\nonumber\\[1mm]
    {\displaystyle\Delta_{n,m}^{l}(x)} &=& 
    {\displaystyle (-1)^{\frac{1}{2}(n+m-l-1)}
    \left(\frac{[n]_x[m]_x[l]_x}{[1]_x}\right)^{\frac{1}{2}}}\\[1mm]
    &\times&{\displaystyle\!\!\!\prod_{j=(l+n-m+1)/2}^{n-1}[j]_x
    \!\!\!\prod_{j=(m+n-l+1)/2}^{n-1}[j]_x
    \!\!\!\prod_{j=(l+m-n+1)/2}^{(l+m+n-1)/2}\frac{1}{[j]_x}}\,,\nonumber
\end{eqnarray}
where the brackets are given by $[j]_x = x^{j/2} - x^{-j/2}$ with
$x = \exp(\pi{\rm i}\alpha_+^2)$ and $x' = \exp(\pi{\rm i}\alpha_-^2)$. 
In the case of interest, $\alpha_+^2 = 2p$ and $\alpha_-^2 = -2/p$.
These formul\ae\ are valid if the
normalization of the two-point functions simply is chosen to be
\begin{equation}
    \Vev{\Psi_{n,n'}(z)\Psi_{m,m'}(w)} =
    \frac{\delta_{n,m}\delta_{n',m'}}{(z-w)^{2h_{n,n'}}}\,,
\end{equation}
where we denote primary fields from the Kac-table by $\Psi$ instead of $\Phi$.
The general normalization constants $N_{(n,n')(m,m')}^{(l,l')} =
\avac{h_{l,l'}}V_{(n,n')(m,m')}^{(l,l')}(1)\vac{h_{m,m'}}$
can be expressed in terms of hypergeometric integrals (the contour
integration of the screening currents around chiral vertex operators of
a free field theory) and are given here for completeness:
\begin{eqnarray}
    \lefteqn{{\displaystyle N_{(n,n')(m,m')}^{(l,l')}\ =\  
      (-1)^{\frac{1}{2}((2n'-1)r+(2n-1)r')}\alpha_{+}^{4rr'}2^{-2rr'}}}
    \nonumber\\[1mm]
    &\phantom{xx}\times& 
    {\displaystyle\prod_{j'=1}^{r'}\frac{[m'-j']_{x'}[j']_{x'}}{[1]_{x'}}
    \prod_{j=1}^{r}\frac{[m-j]_x[j]_x}{[1]_x}}\nonumber\\[1mm]
    &\phantom{xx}\times&
    {\displaystyle\prod_{j'=1}^{r'}\frac{\Gamma(j'\alpha_-^2/2)
    \Gamma(m+(j'-m')\alpha_-^2/2)\Gamma(n+(j'-n')\alpha_-^2/2)}{
    \Gamma(\alpha_-^2/2)
    \Gamma(m+n-2r+(r'-m'-n'+j')\alpha_-^2/2)}}\\[1mm]
    &\phantom{xx}\times&
    {\displaystyle\prod_{j=1}^{r}\frac{\Gamma(j\alpha_+^2/2-r')
    \Gamma(m'-r'+(j-m)\alpha_+^2/2)\Gamma(n'-r'+(j-n)\alpha_+^2/2)}{
    \Gamma(\alpha_+^2/2)
    \Gamma(m'-r'+n'+(r-m-n+j)\alpha_+^2/2)}}\,,\nonumber 
\end{eqnarray}
where $l = n + m - 2r - 1$ and similar for $l'$.
The structure constants of the OPE or equivalently of the Lie-algebra
of the Fourier modes of the chiral fields are then given by
\begin{equation}
    C_{(n,n')(m,m')}^{(l,l')} = {\cal D}_{(n,n')(m,m')}^{(l,l')}
    N_{(n,n')(m,m')}^{(l,l')}\,.
\end{equation}
These expressions have to be evaluated carefully, due to zeroes in
numerator and denominators which arise when $k\alpha_{\pm}^2/2 \in\mathbb{Z}$
for $k\in\mathbb{Z}$. However, it turns out that all these zeroes cancel
nicely, leaving us with well-defined and non-singular structure constants,
when they are considered
as the limit $C_{(n,n')(m,m')}^{(l,l')}(p) = \lim_{\varepsilon\rightarrow 0}
C_{(n,n')(m,m')}^{(l,l')}(p+\varepsilon)$ for $\alpha_+^2/2 = p\in\mathbb{Z}$.
For example, we have for the chiral local primary fields
\begin{eqnarray}
  \lefteqn{\left(C_{(n,1)(m,1)}^{(l,1)}\right)^2\ =\ 
  {\frac{1}{2}(n+m-|n-m|-2)\choose\frac{1}{2}(l-|n-m|-1)}^{-1}
  \varphi(\Delta_{n,m}^{l}(x))^2}\nonumber\\[1mm]
  &\phantom{xx}\times& \prod_{j=1}^{(n+m-l-1)/2}
  \frac{(pj-1)!^2(p(\frac{1}{2}(n+m-l-1)+l+1-j)-2)!^2}{
  (p(m-j)-1)!^2(p(n-j)-1)!^2}\nonumber\\[1mm]
  &\phantom{xx}\times&
  \prod_{j=1}^{n-1}\frac{(p(n-j)-1)!^2}{(pj-1)!(p(n-j+1)-2)!}
  \prod_{j=1}^{m-1}\frac{(p(m-j)-1)!^2}{(pj-1)!(p(m-j+1)-2)!}\nonumber\\[1mm]
  &\phantom{xx}\times&
  \prod_{j=1}^{l-1}\frac{(pj-1)!(p(l-j+1)-2)!}{(p(l-j)-1)!^2}\,,
\end{eqnarray}
where $\varphi(\Delta_{n,m}^{l})$ denotes the phase part of
$\Delta_{n,m}^{l}$, which in this case is just a sign:
\begin{eqnarray}
  &&\Delta_{n,m}^l(x) = (-1)^{lp}(-1)^{\frac{1}{2}(n+m-l-1)(p+1)}
  (-1)^{p((lm+ln+nm)/2-(l^2+m^2+n^2-1)/4)}\nonumber\\[1mm]
  &&\hspace{1cm}\times\frac{\sqrt{nml}(n-1)!^2(\frac{1}{2}(l+m-n-1))!}{
  (\frac{1}{2}(l+n-m-1))!(\frac{1}{2}(m+n-l-1))!(\frac{1}{2}(l+m+n-1))!}\,.
  \phantom{xxxx}
\end{eqnarray}
The other structure constants are not so easily expressed in a closed form
without transcendent functions. Also, some multiplicities have to be taken
into account, which stem from the $SU(2)$ structure of the extended chiral
symmetry algebra. For instance, our fields $\sigma$ and $J$ of the $c=-2$
LCFT are actually spin doublets with respect to this $SU(2)$ structure. 
We present here the complete set of non-trivial OPEs for $c=-2$, i.e.\ $p=2$,
in a graphical form in figure 2 
(omitting the canonical dependencies on the coordinates)
which directly shows the geometrical meaning of them. Thus, one can easily
read off from these graphs that, e.g., the OPE of two twist fields $\mu$
either yields the identity $\mathbb{I}$ or the puncture operator $P$, depending
on whether the two twist fields belong to two different branch cuts or are
joined by a common branch cut.

As is generally the case in LCFTs, highest weight representations are
no longer necessarily irreducible, but may possess a non-trivial
Jordan-cell structure. In the case of the $c_{p,1}$ models, where the
Jordan-cells are of rank two, this results in the following general form
of the OPE:
\begin{eqnarray}\label{eq:myOPE}
    \Psi_{r,s}(z)\Psi_{r',s'}(w) &\sim&\!\!\!
      \sum_{{\scriptstyle R=|r-r'|+1\atop
                \scriptstyle r+r'-1\equiv R\,{\rm mod}\,2}}^{
        r+r'-1}
      \sum_{{\scriptstyle S=|s-s'|+1\atop 
                \scriptstyle s+s'-1\equiv S\,{\rm mod}\,2}}^{
        s+s'-1}\!\!\!\!\!\!\!\!
        C_{(r,s),(r',s')}^{(R,S)}(z-w)^{-h_{r,s}-h_{r',s'}}
\nonumber\\[1mm]
    &\times&\!\!\!\!\!\!\!\!
      \sum_{{\scriptstyle \{k\}\atop
             \scriptstyle |\{k\}|\geq -1-(\Delta h_{R,S}^{})}}\!\!\!\!\!
        a_{(r,s),(r',s')}^{(R,S);\,\{k\}}\!\!
        \left(\partial^{-1-(\Delta h_{R,S}^{})-|\{k\}|}
        (z-w)^{h_{R,S}^{{\rm min}}-1}\right)\nonumber\\
    & &\phantom{xxxxxxx}\times\left({\cal L}_{-\{k\}}\Psi_{R,S}(w)\right)
        \,,
\end{eqnarray}
including only ``descendants'' with respect to the stress energy tensor and
ignoring ``descendants'' with respect to other chiral local
fields such as $J$ and $W$.
Here, ${\cal L}_{-\{k\}}\Psi_{R,S}=L_{-k_1^{}}L_{-k_2^{}}\ldots
L_{-k_n^{}}$ is a level $|\{k\}|=\sum_i k_i$ 
``descendant'' of the primary field $\Psi_{R,S}$, whose conformal
scaling dimension differs by an integer from the one of 
$\Psi_{R,S}$. They form Jordan blocks precisely when $q_{R,S}^{}\in
\mathbb{Z}_+$.
We then have the following relation between the conformal scaling dimensions
of the field $\Psi_{R,S}$ and the lowest primary field of the Jordan
block: $h_{R,S}^{} = h_{R,S}^{{\rm min}} + (\Delta h_{R,S}^{})$ with
$(\Delta h_{R,S}^{})\in\mathbb{Z}_+$. For example, our field $P=\Psi_{1,3}$ has 
$q_{1,3}^{}=1$ but $(\Delta h_{1,3})=0$ such that $(z-w)\partial P=L_0P=
\mathbb{I}$. 
Hence $P$ and $\mathbb{I}$ span the rank two Jordan cell indecomposable highest
weight representation with the following action of the Virasoro zero mode
$L_0$ on it:
\begin{equation}
      L_0|h\rangle = h|h\rangle\,,\ \ \ \ L_0|\tilde{h}\rangle
    = h|\tilde{h}\rangle + |h\rangle\,,
\end{equation}
where in this case $h=0$, i.e.\ $P(z)|0\rangle = |\tilde{0}\rangle$.
This can be seen explicitly from the OPE of two twist fields:
As apparent from the above formula for the OPE,
\begin{eqnarray}
  \mu(z)\mu(w)&\!\sim\!&
    (z-w)^{-\frac14}\!\left[C_{\mu,\mu}^{\mathbb{I}}\mathbb{I} 
    \!+\! C_{\mu,\mu}^P\left(\!P(w)
    \!+\! a_{\mu,\mu}^{P;\,\{0\}}\Big(\partial^{-1}\frac{1}{z\!-\!w}
    \Big)L_0 P(w)\!\right)\!\right]\!+\!\ldots\nonumber\\
    &\!\sim\!& (z-w)^{-\frac14}\left[\mathbb{I} + P(w) - 2\log(z-w)\mathbb{I}\right]
    + \ldots\,,
\end{eqnarray}
producing the desired logarithmic dependency which indicates the 
presence of a non-trivial Jordan cell structure for $P(w)$ and its
``descendant'', the identity field $L_0P(w)=\mathbb{I}\,$.

The coefficients $a_{(r,s),(r',s')}^{(R,S);\,\{k\}}$ can be determined
recursively, since they are entirely fixed by conformal invariance. If
one rewrites the above OPE formula (\ref{eq:myOPE}) by expressing the 
${\cal L}_{-\{k\}}\Phi$ as linear combinations of derivative terms 
$\partial^{\ell}\Phi$, one obtains in the case of {\em chiral local\/} 
fields\footnote[1]{\,(quasi-) primary chiral fields are local iff their 
conformal scaling dimensions $h\in\frac12\mathbb{Z}$.} the
rather simple closed form of the corresponding coefficients
\begin{equation}\label{eq:acoeffs}
    \tilde{a}_{(r,s),(r',s')}^{(R,S);\,\ell} = 
    {h_{r,s}+h_{r',s'}-h_{R,S}-1+\ell\choose\ell}
    {2h_{R,S}-1+\ell\choose\ell}^{-1}
\end{equation}
for $\ell\geq 0$. The general case can be obtained by analytic
continuation, taking into account that the Laurent expansion of a
non-local chiral field receives contributions with fractional powers.
This does not pose a problem in our case, where the fields of the CFT
naturally live on a Riemann surface, i.e.\ on an $p$-sheeted covering
of the complex plane allowing for generalized Laurent expansions with
fractional powers $\alpha\in\frac1p\mathbb{Z}_p$. 
This is the reason for the ``offset'' $\partial^{-1-(\Delta h_{R,S}^{})}
(z-w)^{h_{R,S}^{{\rm min}}-1}$ in (\ref{eq:myOPE}) instead 
of simply $(z-w)^0=1$, on which $\partial^{-|\{k\}|}$ acts. Similar
corrections apply to the generalization of (\ref{eq:acoeffs}).

Moreover, even when $(\Delta h)\not\in\mathbb{Z}$, fields might still be
linked by Jordan-cell structures with respect to other quantum numbers.
In fact, the $c_{p,1}$ LCFTs have the charge $q$ as a further such
quantum number with respect to the current $J$. As one can easily
infer from the form (\ref{eq:vvope}) of the OPE valid for reduced
correlation functions, fields whose charges differ by integers may
also be linked. And indeed, we have in the $c=-2$ case that
the excited twist fields $\sigma,\tau$ are both linked to $\mu$. 
This is the reason, why we had to include $\mu$ and $\sigma$ into
the right hand side of the OPE $\mu\,\nop{J^2(w)}\ =\ \mu(z)W(w)$ as 
generalized ``descendants'' of the field $\tau$. The generalization of
(\ref{eq:myOPE}) for enlarged chiral symmetry algebras is straight forward 
though cumbersome, and is therefore omitted.

The geometrical meaning of the logarithmic divergencies and the Jordan
cell structure in the hyperelliptic case (i.e.\ $p=2$) is that the 
branched covering picture of a
Riemann surface, which intrinsically is not smooth, is ambiguous in the
case of two branch points flowing together. Besides yielding the equivalent
of a puncture, this situation also corresponds to an asymptotic region in
the moduli space of Riemann surfaces, where one handle becomes pinched. 
This latter possibility is accounted for by the logarithmic term in the
OPE of two branch point fields. We have sketched the three different
situations of fusing two branch points in figure 1.
The other possible configurations of the basic fields of the
$c=-2$ logarithmic CFT and their
OPEs (including logarithmic terms) can be inferred in a similar fashion.

When OPEs are regularized by discarding their singular
parts (normal ordering), the logarithmic divergencies will be lost.
In the hyperelliptic case, normal ordering
of two twist fields $\mu$ at a common branch cut
replaces the two confluent branch points 
by a simple pole. Hence, normal ordering can be used to degenerate
the moduli space of, say, $(N_c,N_f)$ supersymmetric
Yang-Mills theories to the one of $(N_c-1,N_f+1)$ theories, thereby
creating a matter hypermultiplet. 
The created hypermultiplets might even be massive, if we
relax the condition that the sum over all branch points must vanish
with the help of a global translation.
Translation is a global conformal transformation and hence satisfies a
Ward identity. However, due to logarithmic divergencies in LCFTs,  
the conformal Ward identity becomes modified,
$L_0\Vev{\ldots}=k\Vev{\ldots}$ with $k$ a non-zero constant.
But accompanying a global translation of all branch points such that
a subset of them sums to zero by appropriate normal ordering of the 
complement subset cancels $k$ and creates massive hypermultiplets instead.

\renewcommand{\theequation}{\mbox{B.\arabic{equation}}}
\setcounter{equation}{0}
\section{Multiple Hypergeometric Functions}

In this Appendix we collect some useful results on the generalized
hypergeometric functions relevant for our purposes. This is mainly the
so-called fourth Lauricella function $F_D$. To begin with, let 
$(a)_n$ be defined for $a\in\mathbb{C}$, $n\in\mathbb{Z}_+$ as
\begin{equation}
    (a)_n = \prod_{k=0}^{n-1}(a+k)\,.
\end{equation}
Of course, this is equivalent to $(a)_n=\Gamma(a+n)/\Gamma(a)$, which
also provides an extension of the definition to arbitrary $n\in\mathbb{Z}$.
Clearly, $(1)_n = n!$ for $n\geq 0$.
In 1893, Lauricella generalized hypergeometric functions to the case
of many variables and defined in particular four multiple series,
$F_A,\ldots,F_D$, which carry his name. The one important to us is $F_D$
which is defined as
\begin{eqnarray}
    \lefteqn{F_D^{(n)}(a,b_1,b_2,\ldots,b_n,c;x_1,x_2,\ldots,x_n)=}\\
    &&\sum_{m_1=0}^{\infty}\sum_{m_2=0}^{\infty}\ldots\sum_{m_n=0}^{\infty}
    \frac{(a)_{m_1+m_2+\ldots+m_n}(b_1)_{m_1}(b_2)_{m_2}\ldots(b_n)_{m_n}}
    {(c)_{m_1+m_2+\ldots+m_n}(1)_{m_1}(1)_{m_2}\ldots(1)_{m_n}}
    x_1^{m_1}x_2^{m_2}\ldots x_n^{m_n}\,,\nonumber
\end{eqnarray}
where, for convergence, we must have $|x_1|,|x_2|,\ldots,|x_n|<1$.
For $n=1$, this function reduces to the ordinary Gauss hypergeometric
function ${}_2F_1(a,b_1;c;x_1)$, and for $n=2$, it is nothing else than
the Appell function $F_1(a;b_1,b_2;c;x_1,x_2)$.

Next, we will provide an integral representation of this Lauricella function
which is of the Euler or Pochhammer type. The point is that it is only
a single integral which means that the function $F_D$ is, in fact, the
general solution of arbitrary $(n+3)$-point correlation functions with at least
one field degenerate of level two. Let us consider the integral
\begin{equation}\label{eq:integral}
    I = \int_0^1u^{a-1}(1-u)^{c-a-1}\prod_{i=1}^n(1-ux_i)^{-b_i}\,{\rm d}u\,.
\end{equation}
Expanding via the generalized binomial theorem, we have
\begin{eqnarray}
 && I = \sum_{m_1=0}^{\infty}\ldots\sum_{m_n=0}^{\infty}\int_0^1
    u^{a-1}(1-u)^{c-a-1}\prod_{i=1}^n\frac{(b_i)_{m_i}}{(1)_{m_i}}
    u^{m_i}x^{m_i}\,{\rm d}u\nonumber\\
    &&=\! \sum_{m_1=0}^{\infty}\!\!\ldots\!\!\sum_{m_n=0}^{\infty}\!
    \frac{(b_1)_{m_1}\!\ldots\!(b_n)_{m_n}}{(1)_{m_1}\!\ldots\!(1)_{m_n}}
    x_1^{m_1}\!\ldots x_n^{m_n}\!\!\int_0^1\!\! u^{a+m_1+\ldots+m_n-1}
    (1-u)^{c-a-1}\,{\rm d}u\nonumber\\ 
    &&= \sum_{m_1=0}^{\infty}\ldots\sum_{m_n=0}^{\infty}
    \frac{(b_1)_{m_1}\ldots(b_n)_{m_n}}{(1)_{m_1}\ldots(1)_{m_n}}
    x_1^{m_1}\ldots x_n^{m_n}\frac{\Gamma(a+m_1+\ldots+m_n)\Gamma(c-a)}
    {\Gamma(c+m_1+\ldots+m_n)}\nonumber\\
    &&= \frac{\Gamma(a)\Gamma(c-a)}{\Gamma(c)}\,
    F_D^{(n)}(a,b_1,\ldots,b_n,c;x_1,\ldots,x_n)\,.
\end{eqnarray}
Indeed, $F_D$ satisfies the following system of partial differential equations
of second order, which can be recognized as the differential equations of
a $(n+3)$-point correlation function containing one field degenerate of
level two:
\begin{equation}
      \Bigg[
    (1\!-\!x_j)\sum_{k=1}^nx_k\frac{\partial^2}{\partial x_k\partial x_j}
    + \left(c\!-\!(a\!+\!b_j\!+\!1)x_j\right)\frac{\partial}{\partial x_j}
    - b_j \sum_{{\scriptstyle k=1\atop\scriptstyle k\neq j}}^n
      x_k\frac{\partial}{\partial x_k} 
    - ab_j\Bigg]F = 0\,, 
\end{equation}
where $F=F_D^{(n)}(a,b_1,\ldots,b_n,c;x_1,\ldots,x_n)$ and $j=1,\ldots,n$.
If hypermultiplets are present, i.e.\ $N_f>0$, then the corresponding
$c=-2$ CFT correlation functions will contain $P(x_j)$ fields which, despite 
of possessing scaling dimension $h(P)=0$, are degenerate of level three. 
Consequently, the reduced correlators
$\Vvevs{\prod_{i=1}^{2g+2}\mu(x_i)
\prod_{j=1}^rJ(x_{2g+2+j})
\prod_{k=1}^sP(x_{2g+2+r+k})}$\linebreak satisfy the following system of
third order differential equations ($n=2g+r+s-1$):
\begin{equation}
    \Bigg[\frac{\partial^2}{\partial x_j^2} - 
  \sum_{{\scriptstyle k=1\atop\scriptstyle k\neq j}}^{n+3}
  \left(\frac{2h_k}{(x_j\!-\!x_k)^2} + \frac{2}{x_j\!-\!x_k}\frac{\partial}{\partial 
  x_k}\right)\Bigg]\frac{\partial}{\partial x_j}\! \prod_{1\leq l<m\leq n+3}
  \!\!\!(x_l-x_m)^{q_lq_m}\Vvevss{\ldots} \!=\! 0
\end{equation}
with $j=2g+2+r,\ldots,n+3$.
Hence, we obtain in particular the following result for an arbitrary period of
an Abelian form on a genus $g$ Riemann surface (the prefactors are
determined via (\ref{eq:trafo2}):
\begin{eqnarray}\lefteqn{
    \Vvevss{\mu(\infty)\mu(0)\mu(1)\mu(x_1)\ldots\mu(x_{2g-1})
    J(z_1)\ldots J(z_r)P(p_1)\ldots P(p_s)\,\nop{P^Q(\varpi)}}}
    \nonumber\\[1mm]
    &=& \pi\prod_{i=1}^{2g-1}(x_i)^{-\frac12}\prod_{j=1}^r(z_j)^1
        \prod_{k=1}^s(p_k)^{-1}(\varpi)^{g+s-r-1}\\[1mm]
    &\times& F_D^{(n+1)}
    (\underbrace{{\textstyle\frac12,\ldots,\frac12}}_{2g\;{\rm times}},
    \underbrace{-1,\ldots,-1}_{r\;{\rm times}},
    \underbrace{1,\ldots,1}_{s\;{\rm times}},
    \underbrace{1+r-g-s}_{=\ 2-\sum_i q_i},
    1;\phantom{xxxxxxxxxxxx}\nonumber\\
    & &\phantom{F_D^{(n+1)}(}
    \frac{1}{x_1},\ldots,\frac{1}{x_{2g-1}},
    \frac{1}{p_1},\ldots,\frac{1}{p_s},\frac{1}{\varpi})\,,\nonumber
\end{eqnarray}
with $n=2g+r+s-1$, $Q=1+r-g-s$ and $\varpi$ 
the image of the zero mode absorbing multi
pole at infinity. In the cases relevant to this paper, the given data
for an arbitrary Riemann surface are its branch points as well as the
poles and zeroes of its metric $\lambda_{\rm SW}$. Therefore, taking
into account charge balance and absorption of zero modes, we find
\begin{eqnarray}\lefteqn{
    \pi^{-1}\Vvevs{\prod_{i=1}^{2g+2}\mu(e_i)\prod_{j=1}^rJ(z_r)
        \prod_{k=1}^sP(p_k)\,\nop{P^Q(\infty)}}}\\[1mm]
    &=& \prod_{i=1}^{2g+2}(\partial_{e_i}M(e_i))^{\frac14}
        \prod_{j=1}^r(\partial_{z_j}M(z_j))^{-\frac12}
        \prod_{k=1}^s(\partial_{p_k}M(p_k))^{\frac12}
        (\infty^2\partial_{\infty}M(\infty))^{\frac12Q}\nonumber\\[1mm]
    &\times& \prod_{i=4}^{2g+2}(M(e_i))^{-\frac12}\prod_{j=1}^r(M(z_j))^1
        \prod_{k=1}^s(M(p_k))^{-1}
        (M(\infty))^{-Q}\nonumber\\[1mm]
    &\times& F_D^{(n+1)}
    \Big(\underbrace{{\textstyle\frac12,\ldots,\frac12}}_{2g\;{\rm times}},
    \underbrace{-1,\ldots,-1}_{r\;{\rm times}},
    \underbrace{1,\ldots,1}_{s\;{\rm times}},
    \underbrace{1+r-g-s}_{=\ 2-\sum_i q_i},
    1;\nonumber\\
    && \phantom{x}
    \frac{1}{M(e_4)},\ldots,\frac{1}{M(e_{2g+2})},
    \frac{1}{M(z_1)},\ldots,\frac{1}{M(z_r)},
    \frac{1}{M(p_1)},\ldots,\frac{1}{M(p_s)},\frac{1}{M(\infty)}\Big)\,,\nonumber
\end{eqnarray}
with again $n=2g+r+s-1$ and $(1)r+(-1)s+(-\frac12)(2g+2)+2$ the total number 
of zero modes. Note the appearance of an additional crossing ratio which
accounts for the zero mode absorbing terms at infinity.

A particular useful identity is the following expansion of $F_D$ in
terms of hypergeometric functions, which can be used to find the
analytic continuations exhibiting logarithmic divergencies:
\begin{eqnarray}
  \lefteqn{F_D^{(n)}(a,b_1,\ldots,b_n,c;x_1,\ldots,x_n)}\nonumber\\[1mm]
    &=&\sum_{m_1=0}^{\infty}\ldots\sum_{m_{k-1}=0}^{\infty}
    \sum_{m_{k+1}=0}^{\infty}\ldots\sum_{m_n=0}^{\infty}
    \frac{(a)_{m_1+\ldots+m_{k-1}+m_{k+1}+\ldots+m_n}
    \prod_{l\neq k}(b_l)_{m_l}}
    {(c)_{m_1+\ldots+m_{k-1}+m_{k+1}+\ldots+m_n}
    \prod_{l\neq k}(1)_{m_l}}\nonumber\\[1mm]
    &\times&\prod_{l\neq k}x_l^{m_l}
    {}_2F_1({\textstyle a+\sum_{l\neq k}m_l,b_k;
    c+\sum_{l\neq k}m_l;x_k})\,,
\end{eqnarray}
valid for $1\leq k\leq n$.
Further information on the Lauricella function $F_D$, in particular
on its analytical continuation properties, can be found, for example,
in \cite{Ext86}.

\end{document}